\documentclass[fleqn,usenatbib]{mnras}

\usepackage{newtxtext,newtxmath}

\usepackage[T1]{fontenc}

\DeclareRobustCommand{\VAN}[3]{#2}
\let\VANthebibliography\thebibliography
\def\thebibliography{\DeclareRobustCommand{\VAN}[3]{##3}\VANthebibliography}

\usepackage{graphicx}	
\usepackage{amsmath}	
\usepackage{float}
\usepackage[T1]{fontenc}
\usepackage{hyperref}
\usepackage[utf8]{inputenc}
\usepackage{tabularx}
\usepackage{soul} 
\usepackage{lineno}

\newcommand{\koral}{\texttt{KORAL}}

\graphicspath{{./}{Figures/}}

\title[Observational Properties of the Puffy Accretion Disks]{Observational properties of puffy disks: radiative GRMHD spectra of mildly sub-Eddington accretion}

\author[M. Wielgus, D. Lan{\v c}ov\'a et al.]{Maciek Wielgus,$^{1,2,3}$\thanks{E-mail: maciek.wielgus@gmail.com (MW)}
Debora Lan{\v c}ov\'a,$^{4,5}$\thanks{E-mail: debora.lancova@fpf.slu.cz (DL)}
Odele Straub,$^{6,7}$
W{\l}odek Klu{\'z}niak,$^{5}$
Ramesh Narayan,$^{3,2}$ \newauthor
David Abarca,$^{5}$
Agata R\'{o}\.{z}a\'{n}ska,$^{5}$
Frederic Vincent,$^8$
Gabriel T\"{o}r\"{o}k,$^{4}$
and Marek Abramowicz$^{4,5,9}$
\\
$^{1}$Max-Planck-Institut f\"ur Radioastronomie, Auf dem H\"ugel 69, D-53121 Bonn, Germany\\
$^{2}$Black Hole Initiative at Harvard University, 20 Garden Street, Cambridge, MA 02138, USA\\
$^{3}$Center for Astrophysics  $|$ Harvard \& Smithsonian, 60 Garden Street, Cambridge, MA 02138, USA\\
$^{4}$Research Center for Computational Physics and Data Processing; Institute of Physics, Silesian University in Opava, Czech Republic\\
$^{5}$Nicolaus Copernicus Astronomical Centre, Polish Academy of Sciences, Bartycka 18, 00-716 Warsaw, Poland\\
$^6$ORIGINS Excellence Cluster, Boltzmannstraße 2, D-85748 Garching, Germany\\
$^7$Max Planck Institute for Extraterrestrial Physics, Gießenbachstraße 1, D-85748 Garching, Germany\\
$^8$LESIA, Observatoire de Paris, Universit\'e PSL, CNRS, Sorbonne Universit\'es, UPMC Univ. Paris 06, Univ. de Paris, Sorbonne Paris Cit\'e, \\ 5 place Jules Janssen, 92195 Meudon, France\\
$^9$Department of Physics, G{\"o}teborg University, SE-412-96 G{\"o}teborg, Sweden
}

\date{Accepted XXX. Received YYY; in original form ZZZ}

\pubyear{2022}

\begin{document}
\label{firstpage}
\pagerange{\pageref{firstpage}--\pageref{lastpage}}
\maketitle

\begin{abstract}
Numerical general relativistic radiative magnetohydrodynamic simulations of accretion disks around a stellar mass black hole with a luminosity above 0.5 of the Eddington value reveal their stratified, elevated vertical structure. We refer to these thermally stable numerical solutions as puffy disks. Above a dense and geometrically thin core of dimensionless thickness $h/r \sim 0.1$, crudely resembling a classic thin accretion disk, a puffed-up, geometrically thick layer of lower density is formed. This puffy layer corresponds to $h/r \sim 1.0$, with a very limited dependence of the dimensionless thickness on the mass accretion rate. We discuss the observational properties of puffy disks, in particular the geometrical obscuration of the inner disk by the elevated puffy region at higher observing inclinations, and collimation of the radiation along the accretion disk spin axis, which may explain the apparent super-Eddington luminosity of some X-ray objects. We also present synthetic spectra of puffy disks, and show that they are qualitatively similar to those of a Comptonized thin disk. We demonstrate that the existing \textsc{xspec} spectral fitting models provide good fits to synthetic observations of puffy disks, but cannot correctly recover the input black hole spin. The puffy region remains optically thick to scattering; in its spectral properties the puffy disk roughly resembles that of a warm corona sandwiching the disk core. We suggest that puffy disks may correspond to X-ray binary systems of luminosities above 0.3 of the Eddington luminosity in the intermediate spectral states.
\end{abstract}




        \section{Introduction}
        \label{intro} 

        Our understanding of accretion onto stellar mass black holes in X-ray binary systems has been built by five decades of observations following the Uhuru satellite mission \citep{Giacconi1971}, and theoretical developments starting with the $\alpha$-disk solution of \citet{Shakura1973}. The former are reviewed in, e.g., \citet{Levine1996,Remillard2006} and the latter in, e.g., \citet{Done2007,accretion_review}. 
        
        Most known black hole X-ray binaries (BHXBs) undergo a cycle of outbursts, driven by a mass accretion rate modulation caused by an outer disk instability \citep{Homan2001, Lasota2001, Baginska2021}. During outburst, in a luminous ($ 0.05  < L/L_{\rm Edd} < 0.3   $) soft (thermal) spectral state, the observations are consistent with multi-blackbody spectra expected from the classic analytic thin $\alpha$-disk model in its relativistic version \citep[NT disk;][]{Novikov1973,Page1974}.
        However, in the radiation pressure dominated regime $p_{\rm rad} \gg p_{\rm gas}$, expected in the inner part of the accretion disk, the analytic model is viscously \citep{viscous74} and thermally \citep{thermal76} unstable. Since the observations indicate stability of BHXB accretion disks on relevant timescales, one must conclude that while the thin disk model is effective in predicting spectra consistent with observations in the applicable luminosity regime \citep{Li2005,McClintock2014}, it \textit{is not} a self-consistent and correct description of the physical reality. For larger mass accretion rates stability can be achieved through dynamical advection, which removes the excess thermal energy, as found for the (vertically integrated) slim disk models \citep[][]{Abramowicz1988,Sadowski2009}, but the puzzle remains for the observationally interesting case of mildly sub-Eddington luminosities. Several stabilizing mechanisms have been suggested \citep[e.g.,][]{Rozanska1999,Oda2009,Ciesielski2012,Zhu2013}, and magnetic fields seem to be particularly promising.
        
        At luminosities above $\sim 0.3\,L_{\rm Edd}$ the observed
        spectra of BHXBs are no longer consistently represented by the thin disk models \citep{McClintock2006}, as the disk geometry begins to deviate from the geometrical thinness assumption. Attempts to utilize the more general theoretical framework of the slim disk model do not fully address the inconsistency \citep{Straub2011}. Unlike thin (or slim) disks, high mass accretion rate flows may in reality include the presence of a warm corona \citep{Zhang2000,Gronki2020}, increasing the system energetic output in soft X-ray, and/or outflows \citep{Ohsuga2009}, modifying the dynamical structure of the system. The impact of the upper disk layers (disk atmosphere) is typically solved for in separation from the radial disk structure \citep[][]{Davis2006,Rozanska2011,Sadowski2011}.  In numerical simulations, the three dimensional structure of the accretion disk can be captured self-consistently.\footnote{Apart from simulations, fully two or three dimensional (analytical) solutions for accretion disks have been obtained only for $\alpha$-disks, e.g., \citet{Kluzniak2000, Regev2002}.}
        
        A numerical model of a magnetically stabilized, radiation pressure dominated, high luminosity 
        accretion disk in the general relativistic radiative magnetohydrodynamic (GRRMHD) framework has been presented by \citet{Sadowski2016}. We have further developed this model in the sub-Eddington case \citep{Lancova2019}, discussing the morphology of the obtained solutions: a dense, geometrically thin disk core is sandwiched by an elevated region of gas with density lower by about $2$ orders of magnitude that nevertheless remains optically thick to scattering. Because of this puffed-up region, supported radially by Keplerian rotation and vertically predominantly by the magnetic pressure, we refer to these solutions as \emph{puffy disks}. 
        
        GRRMHD simulations promise an approach in which the relevant physics (magnetic fields, outflows, cooling, turbulence, etc.) is self-consistently taken into account, at the cost of large computational complexity \citep[e.g.,][]{Ohsuga2011,Sadowski2013, Sadowski2014,Sadowski2017}. In this paper we discuss the observational properties of puffy disks at luminosity of $0.6-1.5$ $L_{\rm Edd}$, inferred from GRRMHD simulations presented in \citet{Sadowski2016} and \citet{Lancova2019}. We 
        discuss how these numerical models could yield a theoretical framework for interpretation of the BHXBs spectra beyond the $\sim 0.3\, L_{\rm Edd}$ thin disk limit.
        



\section{Puffy disks}

\begin{figure}
    \centering
    \includegraphics[width=1.0\columnwidth,trim=0.3cm 0.3cm 0.3cm 0,clip]{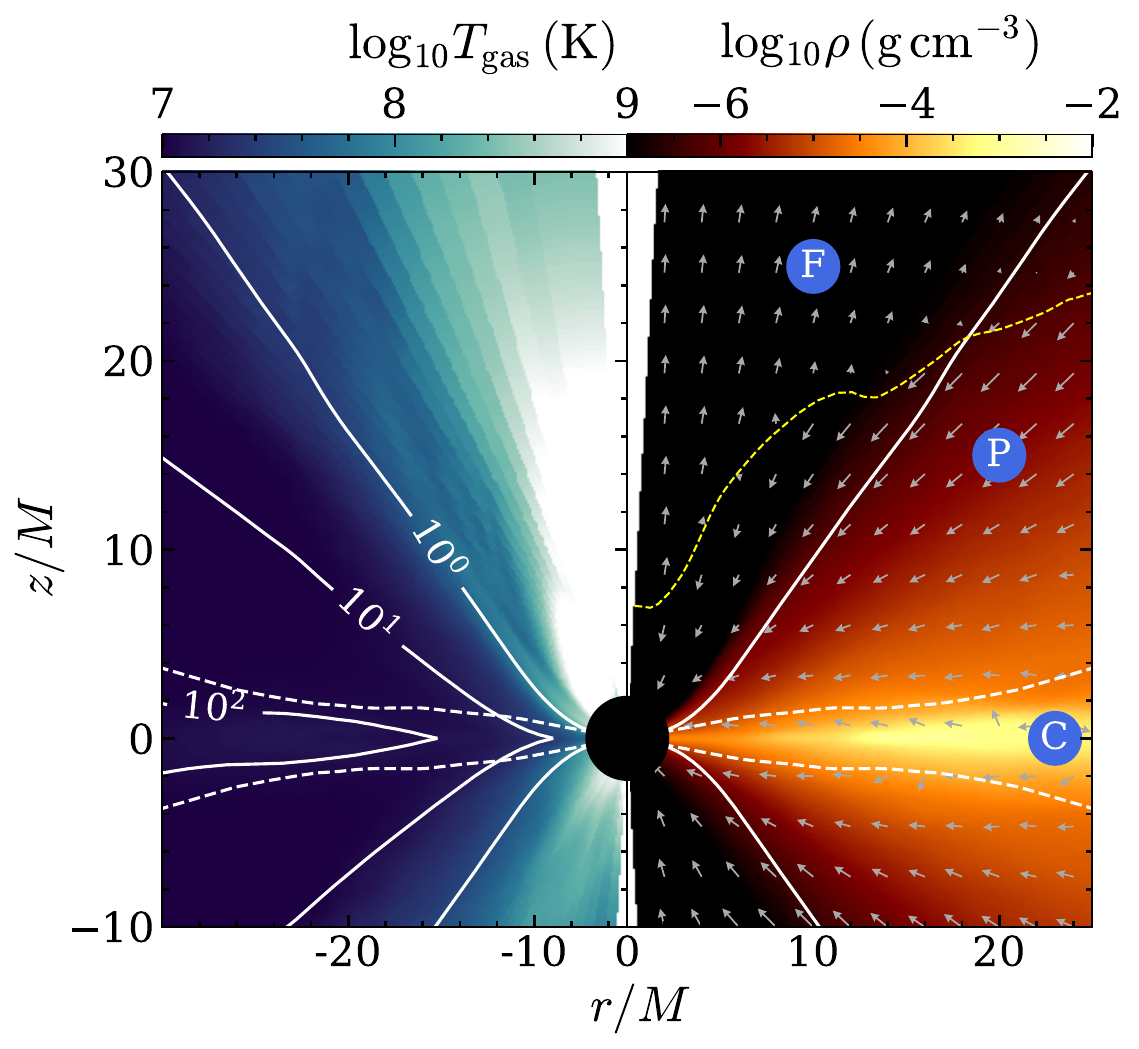}
    \caption{Structure of a puffy disk. The white dashed line in both panels indicates the vertical  thickness (density scale-height) of the dense disk core (region C). \textit{Left:} Gas temperature $T_{\rm{gas}}$ with contours of the optical depth $\tau = 1$ (photosphere, separating regions P and F), $\tau = 10$ and $\tau = 100$ (white continuous lines). \textit{Right:} gas density map, arrows indicate the gas velocity in $r-\theta$ plane. The white contour denotes the position of the photosphere ($\tau\,=\,1$) and the dashed yellow line indicates the border between the inflowing (below) and outflowing (above) regions.}
    \label{fig:puffy_intro}
\end{figure}

\subsection{Numerical setup and conventions}
\label{convent}

We perform GRRMHD simulations of accretion onto a 10\,$M_{\odot}$ Schwarzschild (non-spinning) black hole using \koral,\footnote{\url{https://github.com/achael/koral_lite}} a highly parallelized GRRMHD code \citep{Sadowski2013, Sadowski2014, Sadowski2017}. The details of the numerical setup have been described in \citet{Sadowski2016}. The radiative part of the simulation setup accounts for the electron scattering which dominates the total opacity in the considered regime, as well as free-free absorption and thermal Comptonization \citep[following][]{Sadowski2015}. At this stage frequency-averaged (bolometric) opacities are used, with the aim of capturing the radiation feedback on the disk structure and dynamics. 
We then use the relativistic radiative postprocessing code \texttt{HEROIC} \citep{Zhu2015,Narayan2016} to compute puffy disk spectra and images, solving the complete radiative transfer problem on the converged, time-averaged output of GRRMHD models (gas density, temperature, velocity, and magnetic field maps in the inner part of the simulation volume). The simulations were averaged azimuthally over the whole domain, and in time for a period of at least 5000\,$GM/c^3$, corresponding to a converged, quasi-stationary part of the simulation, mostly unaffected by the initial conditions. The contributions from the outer, unconverged region of the simulation were masked with an extrapolation procedure \citep{Narayan2017} employed for spherical radii $r > 25\,GM/c^2$. \texttt{HEROIC} 
solves for the radiation field in both optically thick and optically thin regimes, including the scattering effects. 

We consider three postprocessed GRRMHD simulation runs, parametrizing the mass accretion rate through the black hole event horizon in Eddington units\footnote{The Eddington luminosity is defined as $L_{\rm Edd} = 4 \pi c \, G M m_{\rm H} / \sigma_{\rm T} = 1.26 \,\times\, 10^{38}\,(M/M_{\odot})$ erg/s, where $G$ is the gravitational constant, $c$ is the speed of light, $M$ is the mass of the gravitating body, $m_{\rm H}$ and $M_{\odot}$ are the hydrogen atom and solar mass, respectively, and $\sigma_{\rm T}$ is the Thompson cross-section.} with $\dot{m}=0.6, 0.9, 1.5$ \citep{Sadowski2016, Lancova2019}. We take
\begin{equation}
    \dot{m} \equiv \frac{\dot{M}}{\dot{M}_{\rm Edd}}   = \frac{0.057 \dot{M} c^2}{{L}_{\rm Edd}} \ ,
\label{eq:mdot}
\end{equation}
where we defined the Eddington mass accretion rate as 
 \begin{equation}
    \dot{M}_{\rm Edd}= {{L}_{\rm Edd} / (\eta c^2)}  = {{L}_{\rm Edd}}/{(0.057 c^2)} \ .
\label{eq:mdotEdd}
\end{equation}
In Eq. \ref{eq:mdotEdd}, $\eta$ denotes the radiative efficiency of accretion, which we set to the thin disk value in the Schwarzschild metric,  $\eta = 0.057$ \citep{Novikov1973}, as a unit convention defining $\dot{M}_{\rm Edd}$.

The actual radiative efficiency in the simulated accretion process, defined through
 \begin{equation}
    \eta= {{L} / (\dot{M}c^2)},
\label{eq:eta}
\end{equation}
can be determined by computing $\dot{M}$ through the horizon and $L$ through a fiducial surface. Note that in presence of advection sphere-integrated luminosity may not be equal to the integrated radiative power produced locally in the disk, causing some additional ambiguity in the efficiency definition. In \citet{Lancova2019} we reported $L = 0.36\,L_{\rm Edd}$ for $\dot{m}=0.6$, measured in the GRRMHD simulation. This implied a significant decrease of radiative efficiency with respect to the NT disk, which we interpreted as a consequence of advection, prominent in the simulation. With the postprocessed radiative transfer we found, somewhat surprisingly, the efficiency of a sub-Eddington puffy disk in the Schwarzschild metric to be consistent with that of the NT disk within the measurement uncertainties. While the discrepancy may be partly explained with the imperfections of the luminosity measurement in the GRRMHD simulation (we integrated the radiative flux over a 25\,$M$ radius sphere in the optically thin region, ignoring contributions from larger radii or optically thick outflows), we currently suspect that this is related to limitations of the M1 closure scheme for the radiation tensor employed in \koral\ \citep{Sadowski2013}. Under the M1 approximation radiation propagates in a single direction in the optically thin regime, and a large part of the radiation released in the inner flow consequently appears to be swallowed by the black hole in the GRRMHD simulation, whereas it may possibly
escape to infinity when a more general radiative transfer problem is solved with \texttt{HEROIC}.
 
\subsection{Properties of puffy disks}

\begin{figure}
    \centering
    \includegraphics[width=\columnwidth]{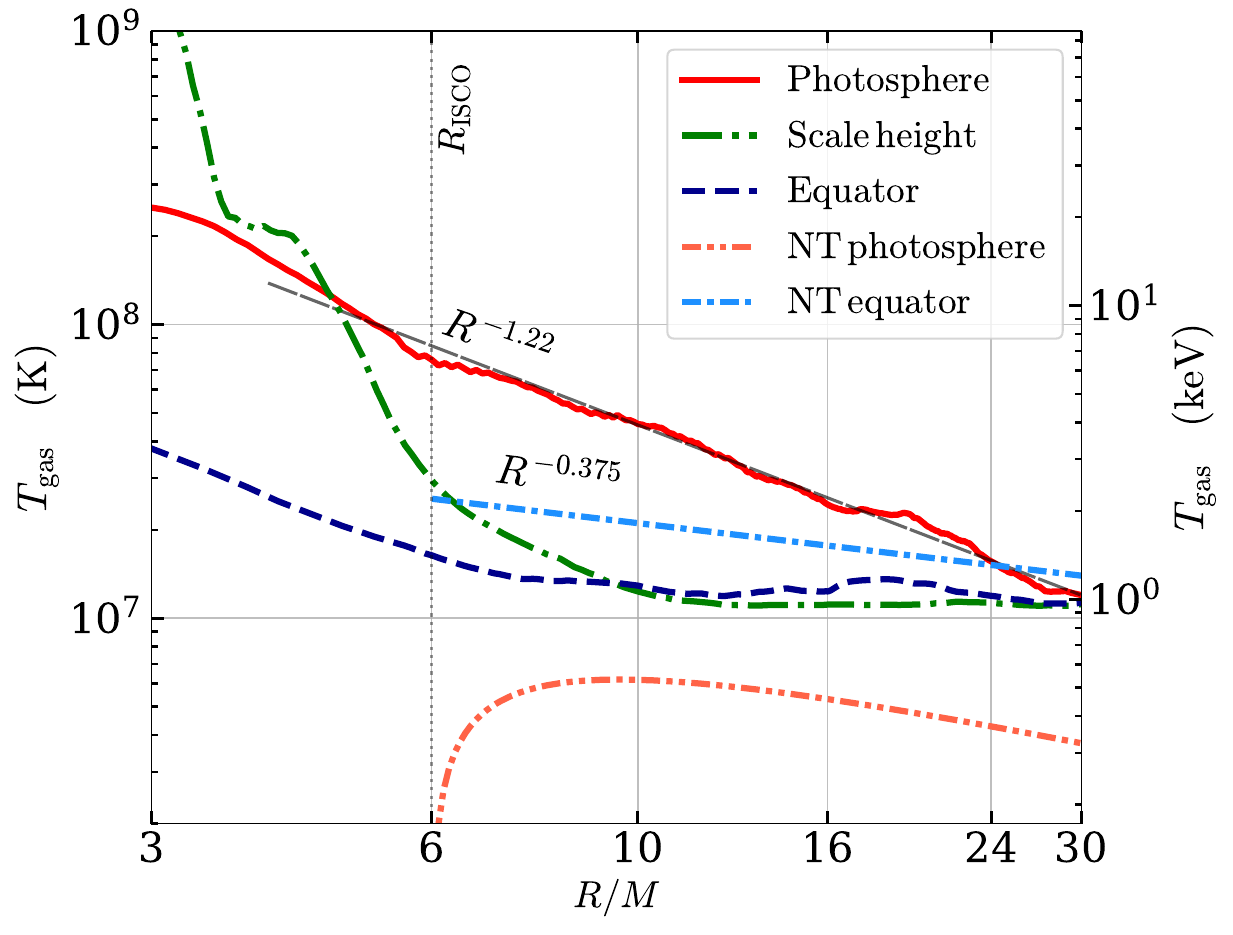}
    \caption{Radial profiles of temperature in the $\dot{m}\,=\,0.6$ puffy disk for the equatorial plane, disk core scale-height, and the photosphere. The photospheric temperature is reasonably well approximated as a power law of the cylindrical radius $R$ with an exponent of $p=-1.22$. The disk core heats up predominantly very close to the black hole, with slow temperature variations for $R > R_{\rm ISCO}$. Analytic NT disk temperature profiles for the same system parameters are shown for comparison.}
    \label{fig:temp_profiles}
\end{figure}

Our numerical simulations predict a stratified structure in a mildly sub-Eddington accretion disk. In Fig.~\ref{fig:puffy_intro} we indicate three zones of the emerging geometry. The zone denoted by C represents the high density core of the disk, with a density scale-height thickness of $h/r \sim 0.1$, which contains most of the gas.
It is supported above the midplane of the disk primarily by magnetic pressure which dominates over radiation pressure, and both strongly dominate over gas pressure \citep{Lancova2019}.
The white dashed line in Fig.~\ref{fig:puffy_intro} shows the density scale height of the disk core. Zone P corresponds to the puffy region. While being optically thick to scattering and delimited by the photosphere at $h/r \sim 1$ (white solid line in the right panel of Fig.~\ref{fig:puffy_intro}), this region is characterized by a gas density that is about 2 orders of magnitude lower than in the disk core. The gas in the puffy region is vigorously accreting onto the black hole. The turbulent dissipation heats up the puffy region and the temperature reaches several times $10^7$\,K at the photosphere, see Fig.~\ref{fig:temp_profiles}. 

In comparing various disk models it is convenient to model the temperature as a power-law function of the cylindrical radius $R$ \citep{Kubota2004}. In the puffy disk, the radial temperature dependence varies from layer to layer in the different zones of the accreting structure. The photospheric temperature scales as $R^{-1.2}$, see Fig.~\ref{fig:temp_profiles}. While this is significantly steeper than for a NT disk, where $T \propto  R^{-0.75}$ \citep{Shakura1973,Novikov1973,Page1974}, the photosphere (surface of last scattering) of a puffy disk is also much hotter in the inner disk region than the NT disk surface. Moreover, the puffy disk photosphere extends below the innermost circular stable orbit (ISCO; $R=6M$ for the Schwarzschild spacetime) with a consistent continuation of the power law temperature dependence, whereas NT surface temperature drops to zero at ISCO. The temperature of the disk core, both in the equatorial plane and at the density scale-height, grows rapidly in the plunging region ($R < R_{\rm ISCO}$) as a consequence of vertical compression, but changes more slowly with radius for $R>R_{\rm ISCO}$. As can be seen in Fig.~\ref{fig:temp_profiles}, these temperature profiles are lower but comparable with the NT disk equatorial temperature $T \propto R^{-0.375}$ for $R > R_{\rm ISCO}$.
At $R > 15M$ optically thick outflows from the outer layer of the puffy zone develop (see the right panel of Fig.~\ref{fig:puffy_intro}). The third zone, denoted by F in Fig.~\ref{fig:puffy_intro} is a low density, low optical depth, hot ($10^8-10^9$\,K) funnel zone. The optically thin material in the funnel, driven by collimated radiation, is outflowing above the stagnation surface located at $z \approx r+7M$.

The morphology of the puffy disk is somewhat similar to the predictions of \citet{Zhang2000} based on  X-ray spectral modelling of BHXBs, and of \citet{Begelman2007}, based on arguments about active galactic nuclei disks.
Conclusions similar to ours about the vertical structure of the disk, including the importance of magnetic pressure have been reached by \citet{Mishra2020}, based on non-relativistic simulations.

\begin{figure*}
\centering
\includegraphics[width=0.85\textwidth]{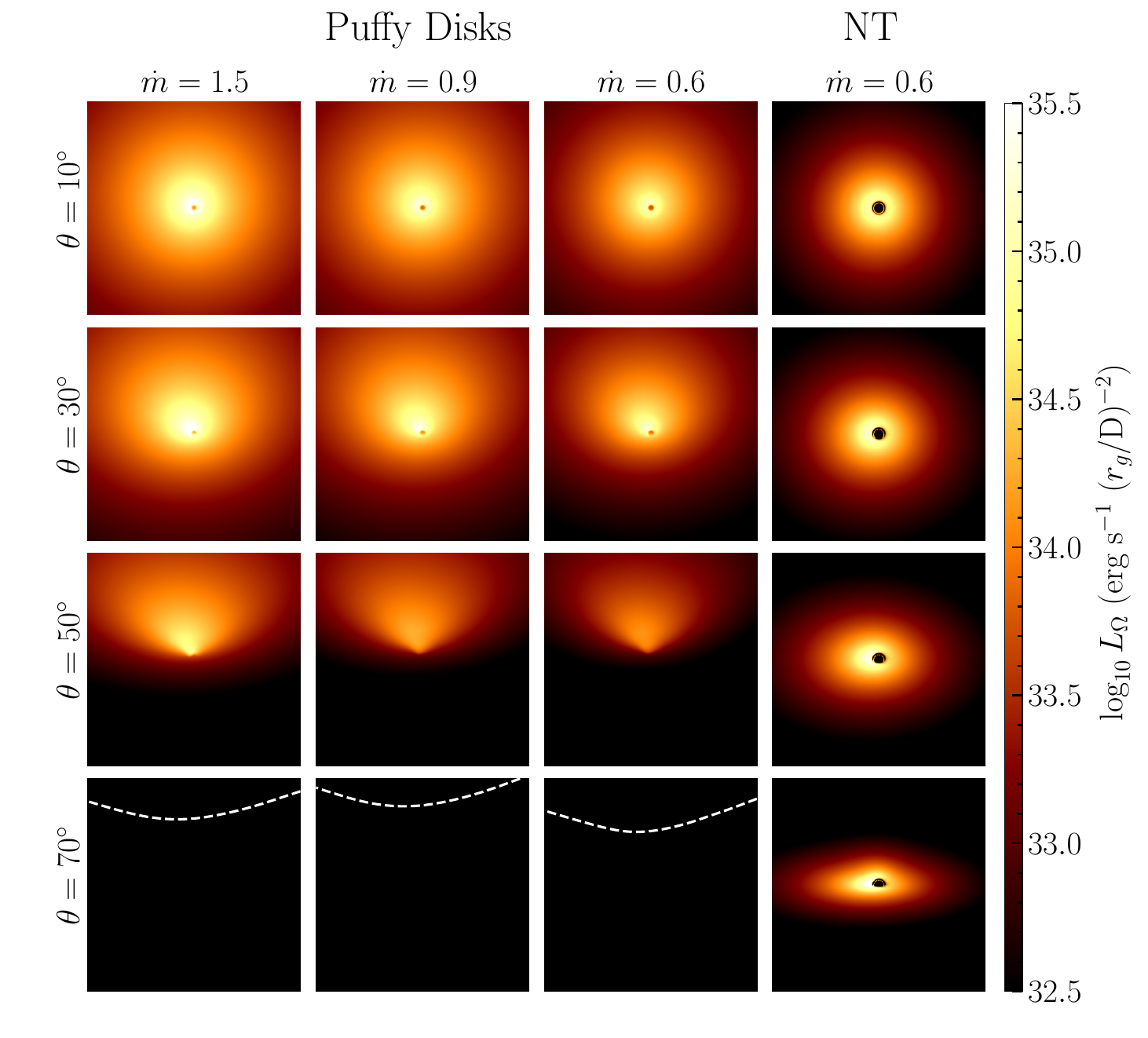}
\caption{Puffy disk appearance given as  $L_{\Omega}$, frequency-integrated isotropic luminosity per solid angle $\left(r_g/D\right)^2$, where $r_g = GM/c^2$ and $D$ represents the observer's distance to the source, using $M = 10 M_{\odot}$, and $D= 10$\,kpc. We show puffy disk images computed for different inclinations $\theta$, for mass accretion rates of 1.5, 0.9 and 0.6 $\dot{M}_{\rm Edd}$ (\textit{first three columns}), compared to the NT disk of 0.6 $\dot{M}_{\rm Edd}$ (\textit{last column}). The white dashed lines in the bottom row of the puffy disk images correspond to $10^{31}$ erg s$^{-1}(r_g/D)^{-2}$ contours -- at the inclination of 70$^\circ$ the inner disk is no longer visible in the images. All results were computed in the Schwarzschild metric and correspond to the $\left(240\,r_g/D\right)^2$ field of view.
		}
\label{fig:appearance}        
\end{figure*}


\section{Beaming and obscuration}
\label{sec:beam_obscur}

The presence of an optically and geometrically thick puffy region has important consequences for the radiative output received by the observer. We illustrate this in Fig.~\ref{fig:appearance}, where ray-traced bolometric (frequency-integrated) images of puffy accretion disks \citep[computed with \texttt{HEROIC}; ][]{Narayan2016} are contrasted with NT thin disk images \citep[computed with \texttt{GYOTO}; ][]{Vincent2011} for several inclinations $\theta$. For the thin disk no vertical structure was assumed. While there seem to be no perspectives in the foreseeable future of observing resolved structures of accretion disks in BHXBs\footnote{The field of view of panels in Fig.~\ref{fig:appearance} corresponds to 2.4\,ns of arc, four orders of magnitude less than the highest resolution achieved by global radio-interferometric arrays \citep{PaperI}.}, this exercise is very helpful in recognizing the influence of inclination and elucidating differences between the NT thin disk  and the GRRMHD puffy disk solutions. At low inclinations ($\theta \le 30^\circ$) observers see the puffy disk system through the low optical depth funnel region. The radiation then becomes collimated in the funnel surrounded by the geometrically and optically thick puffed-up accretion disk in a way similar to the classic analytic thick disk solutions \citep{Abramowicz1978}, and it is enhanced by the reflections from the optically thick funnel walls \citep{Sikora1981}. These effects are self-consistently handled by the \texttt{HEROIC} code. Notice that the mostly purely geometrical beaming effect discussed here is different from relativistic beaming (the Doppler effect related to the emitter approaching with a relativistic velocity), as observed in jets. The latter has relatively little importance in the case of puffy disks, since the outflowing gas inside the funnel has a velocity of only $\sim 0.1\,c$, and a very low density. 

\begin{figure}
    \centering
    \includegraphics[width=\columnwidth]{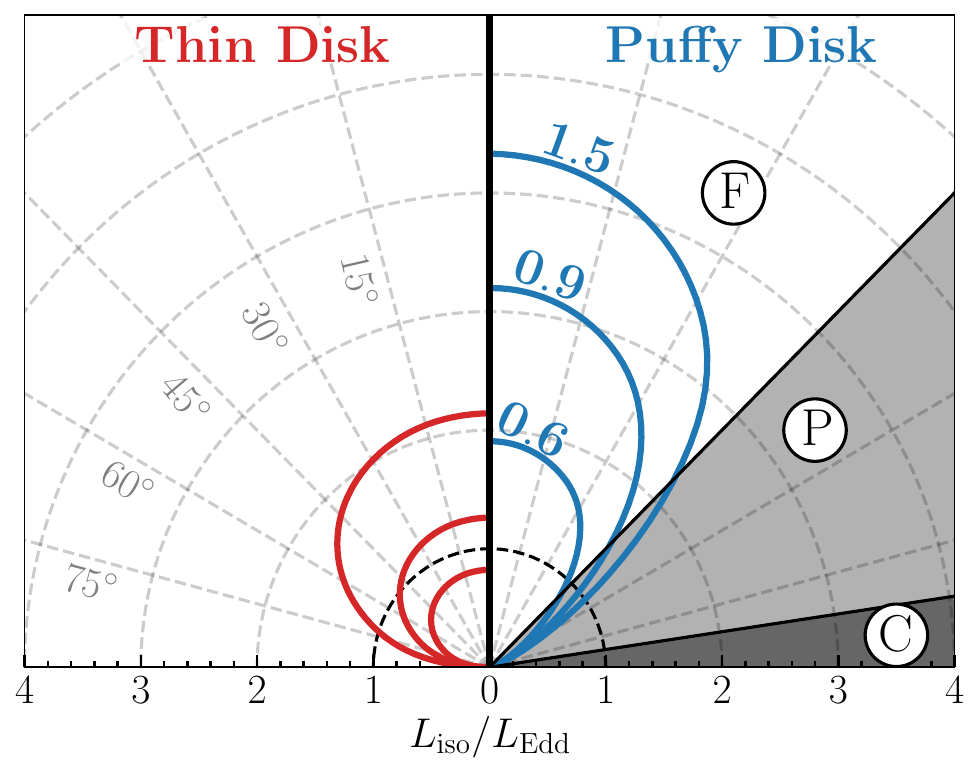}
    \caption{Bolometric beaming pattern comparison between a puffy disk and the NT thin disk, for mass accretion rates of $0.6, 0.9,$ and $1.5 \, \dot{M}_{\rm Edd}$ and spin $a_* = 0$. Puffy disks collimate radiation along the axis of symmetry with a beaming factor $b\approx 0.3$, and the radiation is strongly attenuated for large inclinations.
    }
    \label{fig:beaming}
\end{figure}

The luminosity, $L$, is computed directly from the simulated spectra, at a fiducial distance $D$:
\begin{align}
L &= 2 \pi D^2\int\displaylimits_{0}^{\pi} \left( \int\displaylimits_{0}^{\infty} F_{\nu}\,{\rm d}\nu \right) \sin \theta {\rm d}\theta\\
&= \frac{1}{2}\int\displaylimits_{0}^{\pi} \left( \int\displaylimits_{0}^{\infty} L_{\nu}\,{\rm d}\nu \right) \sin \theta {\rm d}\theta  \ .
\label{eq:luminosity2}   
\end{align}
The collimation of radiation in the funnel causes an increase of the so called isotropic luminosity, $L_{\rm iso}$, that would be inferred from observations at particular inclination $\theta$, with 
\begin{equation}
    L_{\rm iso} = 4 \pi D^2 F  = \int\displaylimits_{\rm image} L_{\Omega}\,{\rm d}\Omega  = \int\displaylimits_{0}^{\infty} L_{\nu}\,{\rm d}\nu \ .
\label{eq:luminosity}
\end{equation}
At the axis ($\theta=0$), the increase corresponds to a factor of $1/b=L_{\rm iso}/L$ relative to the actual luminosity \citep[e.g.,][]{King2001}. In Eq. \ref{eq:luminosity}, $F=\int F_\nu {\rm d}\nu$ is the flux seen by the observer at distance $D$. In fact, the flux $F_\nu$, luminosity per solid angle $L_{\Omega}$, and luminosity per frequency $L_\nu$, all depend on the observer's inclination $\theta$, as can be seen in Fig.~\ref{fig:appearance}, and thus $L_{\rm iso}$ is a function of the same angle (see Fig.~\ref{fig:beaming}). 

 In our simulations we measure a puffy disk beaming factor of $b \approx 0.25 - 0.30$, without a strong indication of dependence on $\dot{m}$, as the general three-zone puffy disk structure does not change much between $\dot{m} = 0.6$ and $\dot{m} = 1.5$. The lack of dependence of the beaming factor on the mass accretion rate contradicts the $b \propto \dot{m}^{-2}$ relation expected from \citet{King2009}, which was derived on the assumption that the disk height increases with the mass accretion rate. We attribute this discrepancy to the roughly constant $h/r$ ratio (regardless of the accretion rate), which is related to advection  limiting the growth of the disk thickness with $\dot{m}$ \citep{Lasota2016,Wielgus2016}. 

At inclination $\theta = 10^\circ$, Fig.~\ref{fig:appearance} shows
that puffy and NT disks are similar in appearance; however, the puffy disk is brighter overall because of the beaming effect. Because of the low optical depth in the funnel, even the innermost region of the accretion flow in the immediate vicinity of the event horizon is visible (the central dark spot in the puffy disk images). This central gap in the image is much more prominent in the NT model, in which radiation is only emitted at radii larger than the ISCO. At larger inclinations a different effect starts to be relevant---the puffed-up optically thick disk starts to obscure the view of the inner part of the accretion flow. In Fig.~\ref{fig:appearance} we see that at an inclination of 50$^\circ$ the puffy disk image is already dominated by the funnel wall opposite to the observer, with the core of the system obscured. At 70$^\circ$ inclination the inner regions of the puffy disk are not visible and the radiative flux decreases by about 2 orders of magnitude. At the same time, the impact of the inclination on the NT disk appearance is much less prominent, as the disk remains geometrically thin, supporting neither efficient beaming nor obscuration effects.

Interestingly, the development of an elevated disk structure during the high mass accretion rate phase, has been proposed as an interpretation of the X-ray spectra of the large inclination source V404 Cygni \citep{Motta2017}. The enhanced complexity of the X-ray outburst light curves of high inclination sources was also identified as hinting at a time-variable obscuration \citep{Narayan2005}. Although low luminosity numerical solutions of stable disk are not yet available, we imagine that a disappearance of the puffy region, and transition between a thin and a puffy disk geometry may occur at some accretion rate, perhaps at $\sim 0.3 \dot{M}_{\rm Edd}$.

We further quantify the collimation / obscuration effects in Fig.~\ref{fig:beaming}, where we compare the beaming pattern of NT thin disks and puffy disks for the mass accretion rates of $\dot{m} = 0.6, 0.9, 1.5$. In general, beaming is maximal at low inclinations and decreases for higher viewing angles and the beaming factor does not strongly depend on the mass accretion rate. The face-on inclination ``beaming'' factor\footnote{In reality, for the thin disk the beaming pattern is related to  the usual non-isotropic emission from the photosphere, the flux decreasing with increasing inclination angle. The unrelated effect of Doppler beaming affects the NT spectra at large viewing angles.}
of the NT disk is $b_{\rm NT} \approx 0.7$ which corresponds to an effect about $2.5$ times less prominent than in the case of the puffy disk. Hence, an NT disk with a mass accretion rate of 1.5\,$\dot{M}_{\rm Edd}$ appears as a 2.1\,$L_{\rm Edd}$ luminosity source when viewed along the axis. A puffy disk of the same mass accretion rate, however, implies an isotropic luminosity of 4.4\,$L_{\rm Edd}$. 

A mildly sub-Eddington puffy disk system viewed at low inclination would be conventionally classified as an ultra-luminous X-ray source \citep[ULX; ][]{Kaaret2017}. We have already noted in Section \ref{convent} that for sub-Eddington accretion, the puffy disk luminosity is proportional to the mass accretion rate, that is, the efficiency does not change appreciably with $\dot{m}$ for $\dot{m} \le 1.5$. Assuming approximate proportionality also between luminosity and black hole mass (with the caveat that, in fact, radiative simulations do not strictly scale with the mass of the black hole), we could give an approximate formula for beamed observed isotropic luminosity of a puffy disk 
\begin{equation}
    L_{\rm iso} \sim 5\times10^{39} \left(\frac{M}{10 M_\odot}\right)\left(\frac{\dot{M}}{\dot{M}_{\rm Edd}} \right)\left(\frac{b}{0.25} \right)^{-1} \frac{\rm erg}{\rm s} .
\end{equation}
This may allow an interpretation of those ULXs that comprise a BHXB in the puffy disk accretion framework.


\begin{figure*}
    \centering
    \includegraphics[width=0.97\textwidth]{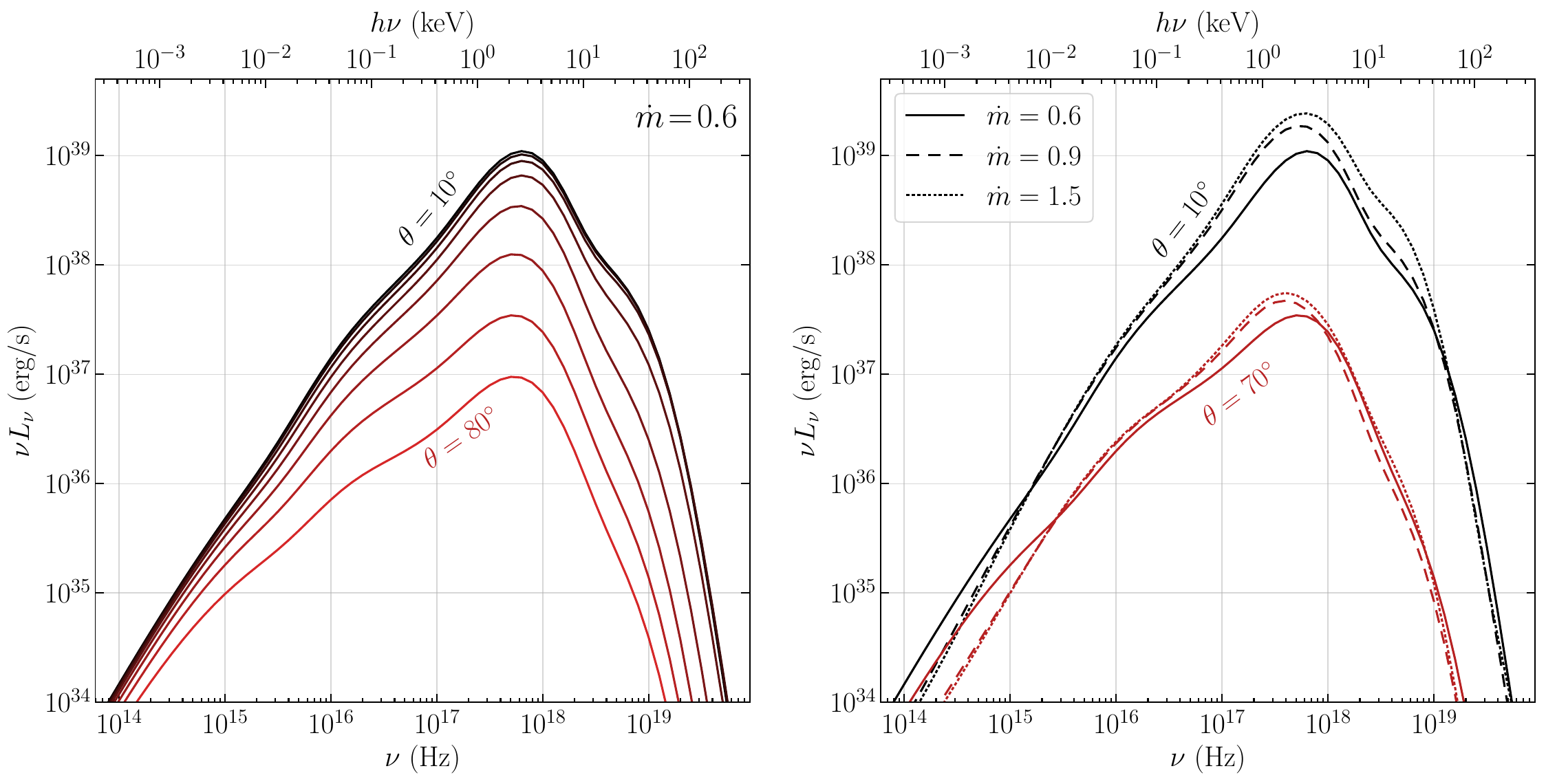}
    \caption{Properties of puffy disk spectra for a $M = 10M_\odot$ black hole.
    \textit{Left:} isotropic radiative power per logarithmic frequency interval, as a function of the observer's inclination, from $10^\circ$ (nearly face-on) to $80^\circ$ (nearly edge on). \textit{Right:} isotropic radiative power per logarithmic frequency interval for three GRRMHD simulations, corresponding to different mass accretion rates $\dot{m}$.}
    \label{fig:spectra_overview}
\end{figure*}

\section{Spectral properties of puffy disks}
\label{sec:spectra}
Traditionally, BHXBs and ULXs are studied with phenomenological models, i.e., by composing a total model from as many components as needed to emulate the observed spectrum: a thin disk, a corona, an iron line, a reflection bump, etc. Given the comparatively simple and analytic nature of the model components, this has the crucial advantage that it allows to identify and characterize different disk properties and effects very efficiently. However, different underlying disk structures necessarily modify the contribution of other components, and this can lead to a misreading of the importance of added effects. Within the GRRMHD framework, we can self-consistently solve for the structure of the accretion system, taking into account the radiation feedback, influence of the magnetic field, vertical stratification, and outflows. All these effects influence the spectra in a highly non-trivial and non-linear way that cannot be fully reproduced by adding and/or multiplying a series of analytic model components. Synthetic spectra from GRRMHD simulations of hard state BHXBs were recently discussed by \citet{Dexter2021}.

\begin{figure}
    \centering
    \includegraphics[width=\columnwidth]{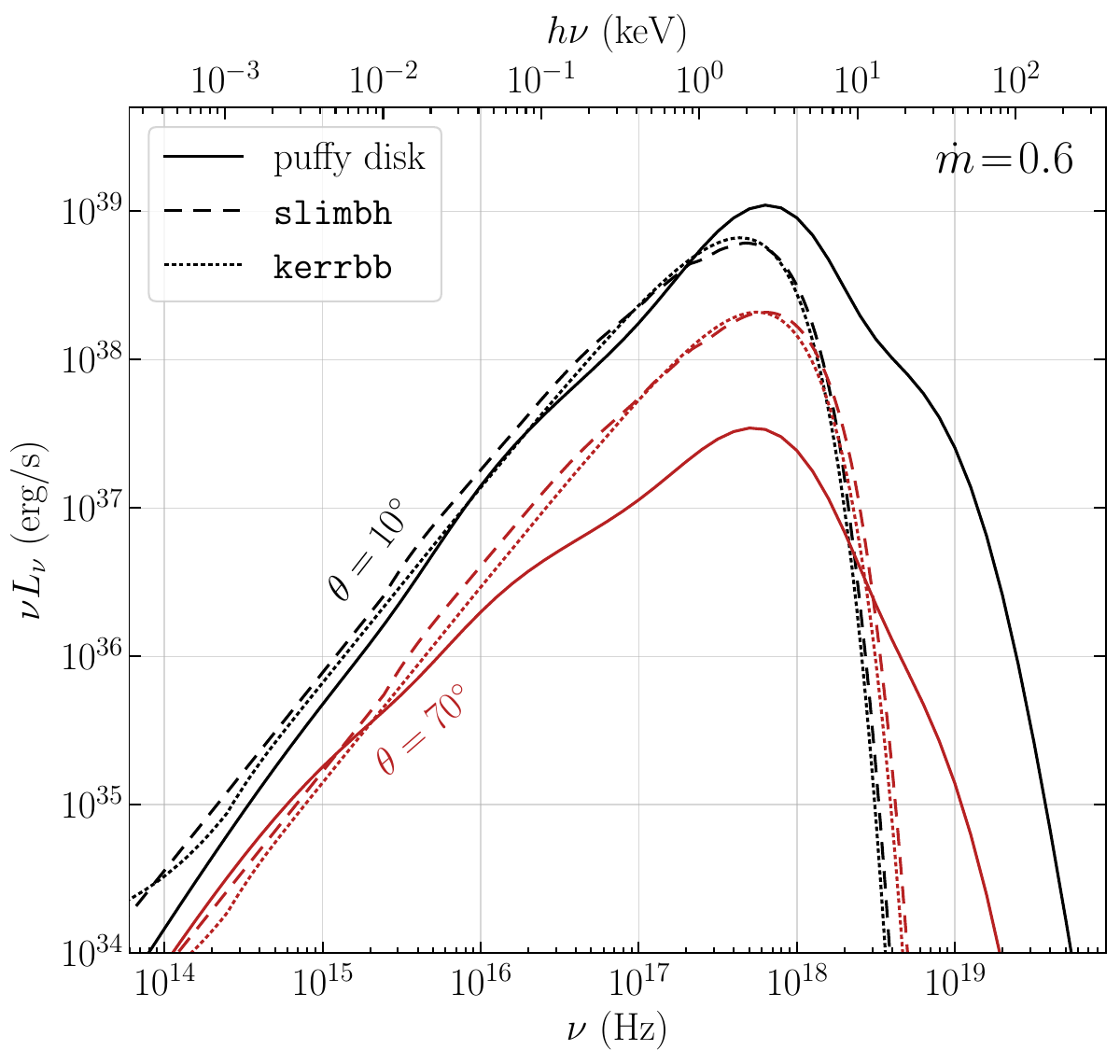}
    \caption{Comparison of puffy disk spectra to multi-color blackbody spectra of thin (\texttt{kerrbb}) and slim (\texttt{slimbh}) disk models of the same mass $M=10M_\odot$, mass accretion rate $\dot{m}=0.6$, and observer's inclination $\theta$.}
    \label{fig:spectra_overview2}
\end{figure}

We present synthetic puffy disk spectra computed using the \texttt{HEROIC} code in Figs.~\ref{fig:spectra_overview}-\ref{fig:spectra_overview2}. In the left panel we explore the influence of the inclination, expanding on the total luminosity discussion from Section~\ref{sec:beam_obscur}. The inclination $\theta$ has a strong impact on the puffy disk spectra because of the beaming at low inclinations, and obscuration at high inclinations. Between $\theta=10^\circ$ and $\theta=80^\circ$ the maximum value of the spectral energy distribution (SED) changes by two orders of magnitude. However, the location of the spectral peak is only mildly affected, and remains around 3\,keV. At low inclinations the SED develops a hump at around 20\,keV, related to the Comptonized radiation produced in the innermost part of the accretion flow.

The right panel of Fig.~\ref{fig:spectra_overview} explores the differences between our three GRRMHD simulations for a range of mass accretion rates and two specific inclinations. Apart from total radiative output scaling, they indicate similar features. A small shift of the peak towards higher energies is visible in the $\dot{m}=0.6$ model. This is related to the lower optical depth of the system.

Finally, in Fig.~\ref{fig:spectra_overview2} we contrast the puffy disk spectra with those of a thin NT disk and a slim disk in their \textsc{xspec} implementations (\texttt{kerrbb} and \texttt{slimbh}, respectively). These are two relativistic, height-integrated $\alpha$-disk models based on \citet[][]{Novikov1973} and \cite{Sadowski2009}, respectively. They both assume a geometrically thin and optically thick disk that consists of a sum of annular black body emitters. This allows one to determine the total disk luminosity by radially integrating $L(r) \propto T^4_{\rm eff}(r)$. Both models parametrize the turbulent viscosity by assuming stress to be proportional to pressure via a constant $\alpha$ \citep{Shakura1973}. The slim disk differs from the thin NT disk on account of its additional advective cooling which transports an increasing fraction of the disk thermal energy towards the black hole as the mass accretion rate rises.

Slim and thin disk spectra appear similar at the moderate accretion rate of $0.6 M_{\rm Edd}$. Neither of them accounts for Compton scattering which can be modelled separately (see Section~\ref{modelling}). This causes a difference with respect to the puffy disk, where Comptonization is self-consistently accounted for. 

\begin{figure}
\centering
\includegraphics[width=0.99\columnwidth]{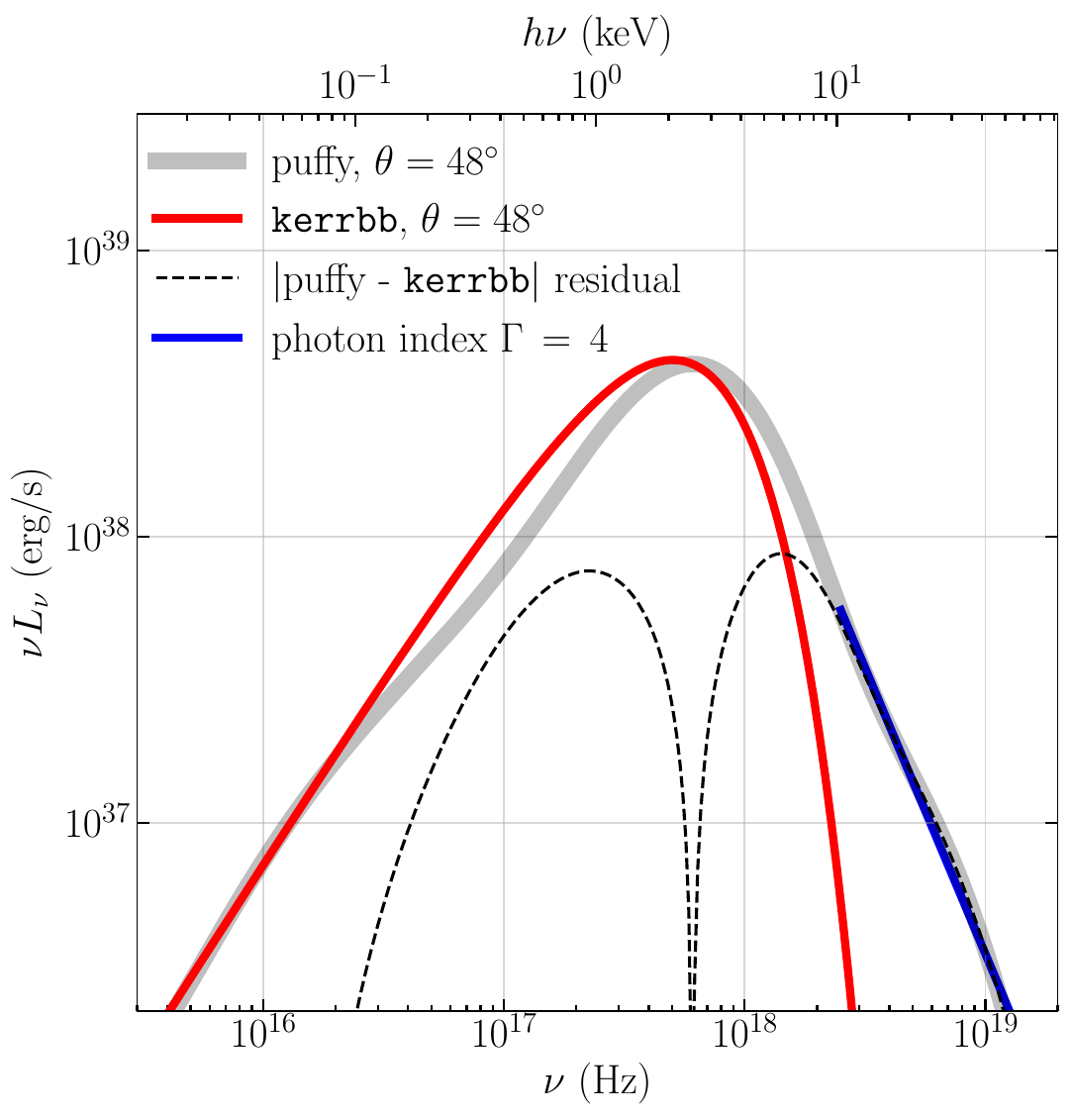}
		\caption{Comparison of NT \texttt{kerrbb} (red lines) and puffy disk (gray lines) spectra at an intermediate inclination of $\theta = 48^\circ$ for a mass accretion rate $\dot{m}=0.6$. The absolute value of the difference between the models is shown by dashed lines. In the range of 10-40\,keV the puffy disk spectrum can be approximated with a power law corresponding to a photon index $\Gamma=4$ (blue line).
		}
\label{fig:puffy_kerrbb}        
\end{figure}
\begin{figure}[b]
\centering
\includegraphics[width=0.99\columnwidth]{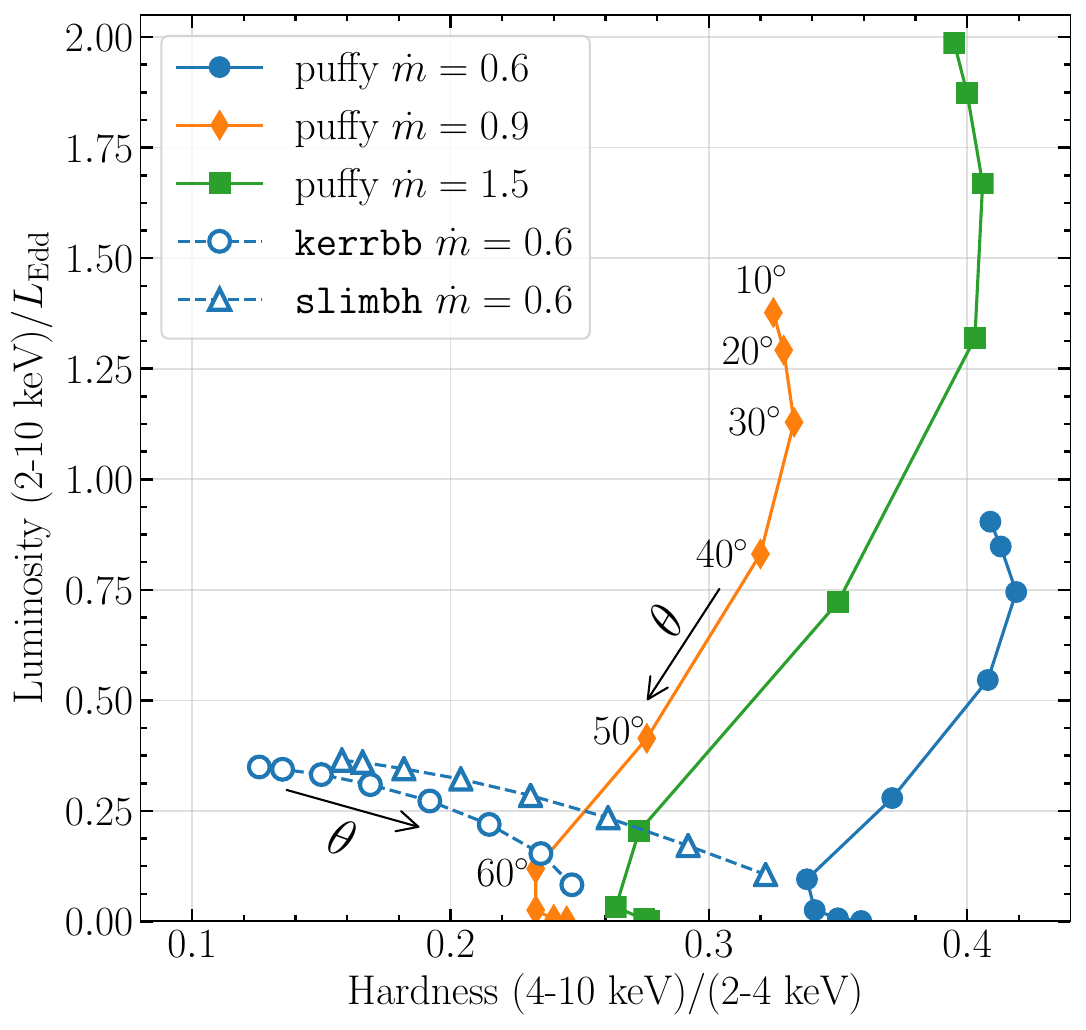}
		\caption{Luminosity-hardness diagram for the synthetic puffy disk spectra. Vertical axis represents the 2-10\,keV band isotropic luminosity in the Eddington units. Hardness is defined as the ratio of photon counts in the 4-10 keV and 2-4 keV bands. For comparison, the location of the \texttt{kerrbb} and \texttt{slimbh} spectra for $\dot{m} = 0.6$ are shown as well; for small and moderate inclinations they both indicate less luminous and softer spectra than a puffy disk at the same mass accretion rate.
		}
\label{fig:lumin_hardn}        
\end{figure}

For intermediate inclinations, around $\theta=48^\circ$, where neither beaming nor obscuration are dominant, the puffy disk peak
of SED is similar to that of the NT disk model. The comparison between NT and puffy disk spectra at $\theta=48^\circ$ is shown in Fig.~\ref{fig:puffy_kerrbb}. There is a deficit of power in the puffy spectrum with respect to the NT disk in the range of 0.1-1.0\,keV and an excess of puffy disk power above 3\,keV. The puffy disk spectrum above 10\,keV is described reasonably well by a power law with a $\nu L_\nu$ slope equal to $-2$, corresponding to a photon index $\Gamma = 4$.

The effect of inclination on the luminosity and hardness of the spectra for different disk models is shown in Fig.~\ref{fig:lumin_hardn}. As discussed in Section~\ref{sec:beam_obscur}, the observer's inclination has a huge impact on the observed puffy disk luminosity. The hardness of puffy disk spectra diminishes between $30^\circ$-60$^\circ$ inclination because of the obscuration of the inner (beamed hard X-ray) region of the disk. In comparison, both the thin and the slim disk are very soft at low inclinations as they completely lack the hard Compton component. They become harder under increasing viewing angles due to relativistic beaming in directions close to the plane of the disk, primarily affecting the inner region of the disk \citep[e.g.,][]{Kulkarni2011}.
Moreover, at $\dot{m}=0.6$ slim disk spectra tend to be harder than those of an NT thin disk, owing to advection that traps thermal energy in the disk, transports it inwards and releases it at smaller radii as harder emission. The non-monotonic relation between the spectral hardness and mass accretion rate for the puffy disks, seen in Fig.~\ref{fig:lumin_hardn}, is more difficult to understand and can be attributed to an interplay between the temperature of the disk, which depends on the mass accretion rate, and the obscuration.



\section{Modelling puffy disk spectra}
\label{modelling}

\begin{figure*}
\centering
\includegraphics[width=0.99\textwidth]{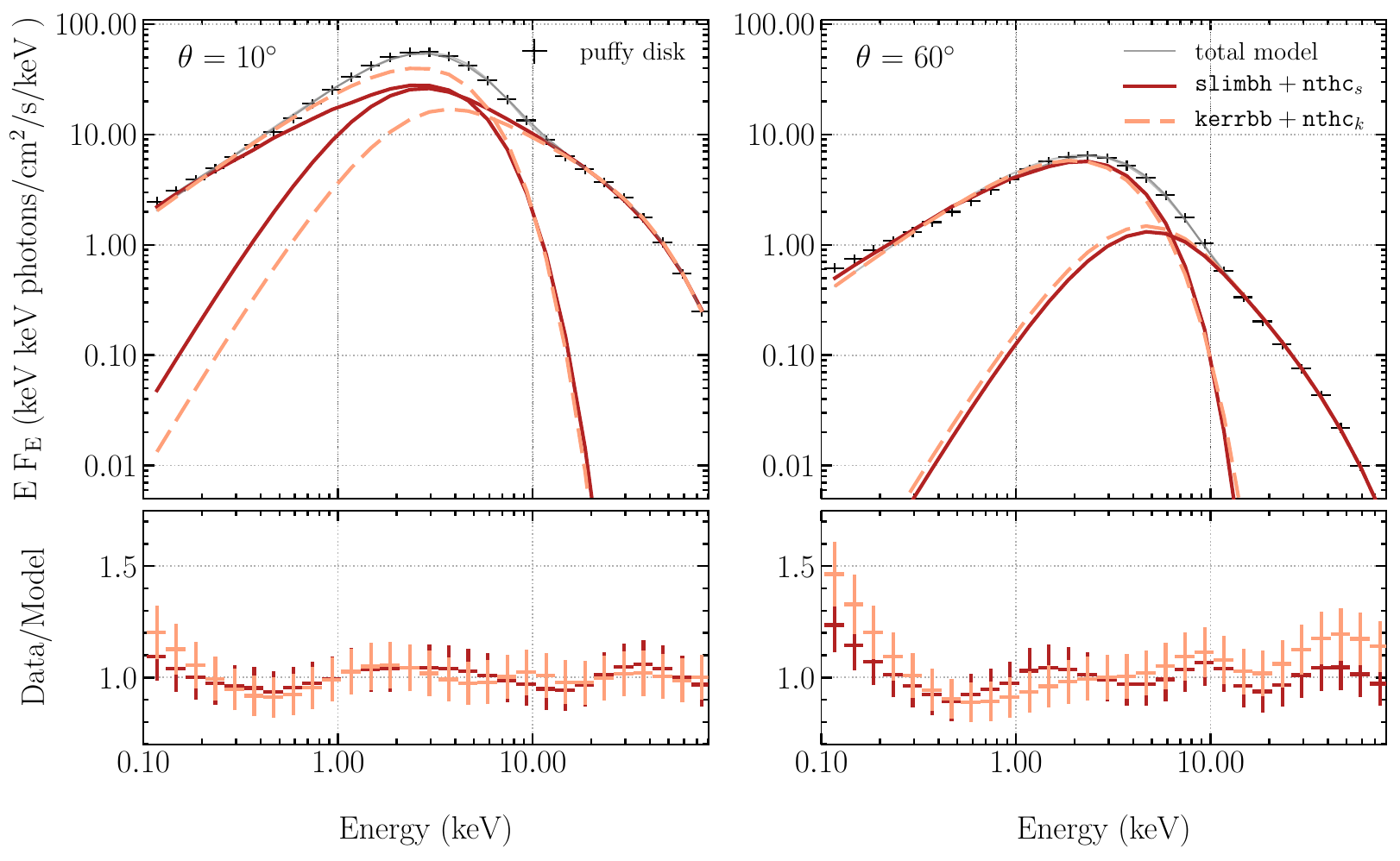}
		\caption{
		Synthetic puffy disk spectra fitted by standard \textsc{xspec} models. The synthetic puffy disk spectra (black crosses) are calculated for a hypothetical BHXB at $D = 10$\,kpc, with a black hole mass $M = 10 M_{\odot}$, a mass accretion rate $\dot{m} = 0.6$, and inclinations $\theta = 10^{\circ}$ (left panel) and $\theta = 60^{\circ}$ (right panel). They are fitted with a disk model plus a Compton component: \texttt{slimbh}+\texttt{nthcomp}$_s$ (solid lines) and \texttt{kerrbb}+\texttt{nthcomp}$_k$ (dashed lines).
		}
\label{fig:puffyfits}        
\end{figure*}

During their outbursts, X-ray binaries pass a series of different disk states and typically spend a comparatively long time in a highly luminous, disk dominated state with a characteristic black body dominated SED. Strictly thermal dominated spectra, i.e., spectra with a disk-to-total flux fraction of above 75\% (in the energy range 2-20\,keV), a weak non-thermal power continuum, and very weak variability (rms < 1\%), can be used to extract the black hole spin \citep[continuum fitting method, e.g.,][]{Remillard2006,Done2007,McClintock2014}. 

If we were to observe spectra from an astrophysical object consistent with our puffy disk model, how would they be interpreted by the standard spectral fitting tools? To address this question we generated synthetic test spectra from a puffy disk simulation in the energy range 0.1-100\,keV, and fitted them with the \textsc{xspec} software package \citep{Arnaud1996}. The synthetic puffy spectra correspond to a GRRMHD simulation of a Schwarzschild black hole of spin $a_* = 0$, mass accretion rate $\dot{m} = 0.6$ (as defined in Eq.\,\ref{eq:mdot}), and mass $M = 10 M_{\odot}$, at a distance D = 10\,kpc, seen from two different inclinations, $\theta = 10^{\circ}$ and $\theta = 60^{\circ}$. We fit these 
spectra with a fully relativistic (including gravitational bending of photon trajectories) thin disk model, \texttt{kerrbb} \citep{Li2005}, and a fully relativistic, advective slim disk model, \texttt{slimbh} \citep{Sadowski2009, Straub2011}. Each disk component is accompanied by a thermal Comptonization component \texttt{nthcomp} \citep{Zdziarski1996,Zycki1999} which accounts for seed photons originating in the disk that are upscattered to higher energies by a homogeneous corona.

For each disk model the dynamical BHXB parameters, mass $M$, distance $D$, and inclination $\theta$, are assumed to be known and fixed, while the black hole spin $a_*$ and total (spherically-integrated, inclination-independent) disk luminosity $L_{\rm disk}$ (in \texttt{kerrbb}, equivalently, the mass accretion rate) are fitted for. The viscosity parameter is fixed at $\alpha=0.1$ in each model. In the Compton component, the temperature of the seed photons, $T_{\rm bb}$, and the electrons in the corona, $T_{\rm e}$, as well as the photon index, $\Gamma$, are fitted for. Usually, in order to reduce the number of free parameters, $T_{\rm bb}$ is fixed at some ``reasonable'' value. Here, we leave it free to assess how well the $T_{\rm bb}$ parameter matches the puffy disk temperature, as presented in Fig.~\ref{fig:temp_profiles}.

\begin{table}
\begin{center}
\vspace{0.5cm}
\tabcolsep=0.1cm
\renewcommand{\arraystretch}{1.35}
\begin{tabularx}{0.84\linewidth}{ l c  c | c  c }      
\hline\hline

& \multicolumn{2}{c}{puffy $\theta=10^\circ$} &  \multicolumn{2}{c}{puffy $\theta=60^\circ$}\\
       &  \texttt{slimbh}  &\texttt{kerrbb} & \texttt{slimbh} & \texttt{kerrbb} \\ 
\hline
$a_*$              & $0.82^{+0.08}_{-0.16}$ & $0.86^{+0.08}_{-0.14}$  & $0.55^{+0.11}_{-0.29}$  & $0.48^{+0.12}_{-0.41}$   \\
$L_{\rm disk}/L_{\rm Edd}$    & $0.73^{+0.15}_{-0.16}$ & $0.79^{+0.1}_{-0.1}$  & $0.21^{+0.19}_{-0.24}$  & $0.17^{+0.05}_{-0.03}$      \\
$\Gamma$           & $2.85^{+0.17}_{-0.20}$ & $2.72^{+0.24}_{-0.24}$      & $3.62^{+0.27}_{-0.29}$  & $3.67^{+0.34}_{-0.27}$  \\
$T_{\rm e}$\,(keV)  & $14.95^{+5.31}_{-2.68}$& $13.05^{+4.62}_{-2.43}$  & $19.51^{+17.34}_{-5.75}$& $19.26^{+21.81}_{-5.58}$  \\    
$T_{\rm bb}$\,(keV) & $0.50^{+0.08}_{-0.07}$ & $0.67^{+0.22}_{-0.15}$  & $1.01^{+0.57}_{-0.31}$   & $0.97^{+0.48}_{-0.29}$   \\       
$\chi^2/dof$       & 9.32/29  & 8.51/29  & 11.58/29  & 27.97/29\\ 
\hline\hline
\end{tabularx}{}
\end{center}
\vspace{-0.25cm}
\caption{\textsc{xspec} fits to synthetic puffy disk spectra with $a_* = 0$ and $\dot m =0.6$.
Each of the disk models, \texttt{slimbh} and \texttt{kerrbb}, is accompanied by the thermal Comptonization model \texttt{nthcomp}.}
\label{tab:fits} 
\end{table}

Given the phenomenological nature of the models, they are both able to fit the puffy disk spectra with high fidelity (low $\chi^2$ per degree of freedom, $\chi^2/dof$), as shown in Fig.~\ref{fig:puffyfits} and Tab.~\ref{tab:fits}. We assumed a 10\% error budget on the simulated spectral measurements. While the fits are well behaved, the models formally overfit our simulated data due to the assumed uncertainties. Nevertheless, the fits demonstrate that all quantities can be well constrained. Furthermore, typical values for BHXBs are obtained, albeit with results that are quite different from the true values of $a_* = 0$ and $\dot m= 0.6$.

The highly beamed low inclination ($\theta = 10^\circ$) puffy spectrum is preferentially fitted with a high black hole spin $a_* \approx 0.7 - 0.8$ for both disk models, and a (thermal) disk luminosity that is slightly higher than the true luminosity of the puffy disk, but much lower than its isotropic luminosity at this inclination angle. The Comptonization component fits a photon index of about $\Gamma$ =\,2.8, a disk temperature of $T_{\rm bb} \approx 0.6\,$keV, and an electron temperature of approximately $T_{\rm e}$ = 14\,keV. While the thin and slim disk models require a cranked up spin parameter to fit the low-energy part of puffy spectra, the Compton model recovers the shape of the high-energy emission with typical values for a warm corona \citep[see, e.g.,][]{Rozanska2015}. The dim, inclined puffy disk ($\theta = 60^\circ$) is fitted with moderate black hole spins and very low disk luminosities, combined with a warm corona of $\Gamma \approx 3.6$, $T_{\rm bb} \approx$ 1\,keV, and $T_{\rm e} \approx 19$\,keV.

The measurement of spin is impaired by the large Compton contributions, which is a known obstacle to the continuum fitting method and one of the reasons why only the strictly thermal dominated states can be effectively used to measure black hole spins, as pointed out by, e.g., \citet{Gierlinski2004,Done2006,McClintock2006,Middleton2006}. Only limited attempts have been made to extend the continuum fitting method towards modelling spectra with an appreciable power law component, e.g., \citet{Steiner2009}.

In Tab.~\ref{tab:fits} we show the parameters of the best fits with uncertainties corresponding to 90\% confidence intervals. The classical disk models gravely overestimate the black hole spin, with respect to $a_* = 0$ that was assumed in the underlying GRRMHD simulation. This overestimate reflects the high temperatures in the puffy disk, seen in Fig.~\ref{fig:temp_profiles}. The mass accretion of the fitted models correspond to $\dot{m} \approx 0.4$ for low inclination and $\dot{m} \approx 0.1$ for large inclination (with $\dot{m}$ defined using the zero-spin efficiency convention from Eq.~\ref{eq:mdot}), i.e., the fitted $\dot M$ value is only equal to 2/3 and 1/6, respectively, of the true value in the simulation. The mismatch with respect to the true value of $\dot{m} = 0.6$ is explained by the increased disk efficiency at high black hole spin, while the variation of the fitted mass accretion rate with the inclination angle involves beaming/obscuration effects as well.


Puffy disk spectra do not describe the strictly thermal dominated state in BHXBs; they are not strongly dominated by thermal radiation but comprise of a considerable amount of Compton up-scattered photons (see Fig.~\ref{fig:spectra_overview}). Their global (0.1-100 keV) blackbody-to-total flux ratio, as measured by \textsc{xspec}'s thin and slim disk models, lies for $\theta=10^{\circ}$ at 54\% and for $\theta=60^{\circ}$ at 83\%, whereas in the 2-20 keV range the disk contribution is just 44\% and 72\%, respectively. In combination with their rather steep high-energy slopes (see $\Gamma$ in Tab.~\ref{tab:fits}) and high normalizations these properties can be related to the spectra observed in two accretion disk states. On the one hand they resemble the so-called intermediate states in BHXBs. These are complex states associated to transitions between the hard and the soft state and the occurrence of quasi-periodic variability \citep[][]{Mendez1997,Remillard2006}.  Although puffy disks can exhibit, depending on the viewing angle, a considerably larger (even super-Eddington) apparent luminosity than average BHXBs, a spectral resemblance is expected due to the presence of a Comptonized corona. On the other hand, puffy disk spectra show morphological similarities to certain ULXs, namely those that show a soft ultraluminous state. There is increasing evidence that at least some ULXs are powered by accreting stellar mass black holes that produce three specific types of ultraluminous disk states: a single-peaked broadened ultra-soft disk, a two-component hard ultraluminous, and a soft ultraluminous state. The latter is typically described as a cool disk with a warm corona component \citep[see][]{Gladstone2009, Sutton2013, Mondal2021}, which resembles our puffy disk spectra.


\section{Discussion}
\label{discuss}

In this paper we discussed the observational signatures, and in particular the spectra, of the puffy disk---a numerical model of stable accretion in the range of luminosities (currently) above $0.5 L_{\rm Edd}$. Unlike \citet{Narayan2017} or \citet{Ogawa2021}, who treated a strongly super-Eddington case, we focused on mildly sub-Eddington accretion. This is the regime where standard geometrically thin disk models fail to correctly predict the observed source spectra, which happens at luminosities above $0.3 L_{\rm Edd}$. The puffy disk solution captures the physics of such luminous accretion flows. In our approach, radiation is incorporated into the GRMHD simulations, allowing for a self-consistent development of a vertically stratified, magnetic and radiation pressure dominated, accretion structure. A low density ``puffy'' region, Keplerian, rapidly accreting, and geometrically and optically thick,  develops above the denser disk core zone; spectroscopically, it resembles a ``warm corona''. 

The observable luminosity of puffy disks depends much more strongly on the observer's inclination to the system axis than is the case for standard geometrically thin disk models. At high inclinations the emission is obscured by the geometrically thick corona blocking the view of the central region of the disk. The development of an obscuring geometrically thick structure during the luminous thermal state may explain unusual variability of BHXB sources, e.g., V404 Cygni. At low inclinations the emission is geometrically collimated by the puffy region, increasing the observed radiative flux. Such a strong  beaming effect is absent in thin disks, and may be relevant for the interpretation of at least some ULX sources. 

We investigated synthetic spectra of puffy disks and attempted to fit them with standard fitting models available in \textsc{xspec}. We found that a comptonized NT disk fits the puffy disk spectra well in terms of the statistical fit quality metrics. However, for our model of a non-spinning black hole the currently available fitting tools erroneously find a large positive spin value. This offers a theoretical interpretation for known problems with fitting BHXB spectra in luminous states, although we can not currently provide a quantitative characterization of these effects in the Kerr metric. For instance, for LMC X-3 \citet{Straub2011} and \citet{Steiner2014} reported that the spin estimate of $a_* \approx 0.21$ decreases with luminosity due to the incompleteness of physical models (lack of Comptonization) at luminosities larger than about 0.3 $L_{\rm Edd}$. For the Schwarzschild puffy disk solution we can only see an increase of the spin estimate with respect to the ground truth value of $a_* = 0$, and indeed this is what happens as the standard \textsc{xspec} models try to emulate the puffy disk spectra by tuning the free parameters.

The puffy spectra resemble intermediate states between soft and hard emission states of BHXBs. The complex intermediate states are associated with quasi-periodic variability and there currently exists no established accretion disk model to describe these states. We have demonstrated that numerically calculated puffy disk spectra, which self-consistently model accretion disks with a significant Comptonized atmosphere (warm corona), could qualitatively describe these intermediate states. It remains to be seen if GRRMHD framework could provide a quantitatively correct interpretation of the real astrophysical sources, extending the region of applicability of the continuum fitting spin estimation method. 

It is currently computationally too expensive to produce a fine-mesh grid in the parameter space of puffy disks, to calculate a table model that could be implemented for public use in the \textsc{xspec} software package. For this reason, we cannot at present explicitly fit puffy disk models to observational X-ray spectra. However, this is clearly one of the directions that should be pursued in the future. Currently, we do not have puffy disk solutions for the Kerr metric with $a_*\ne 0$, nor for $L<0.6L_{\rm Edd}$. The latter solutions are urgently needed, as there as yet exist no known stable solutions for disk accretion at rates corresponding to the luminosity of many BHXB sources.

\section*{Acknowledgements}

The authors thank Andrew Chael, Brandon Curd, and Aleksander S{\k a}dowski for support with the \koral\ code, as well as Silke Britzen, Michal Bursa, and Monika Mo{\'{s}}cibrodzka for valuable comments and discussions. The computations in this work were supported by the \textsc{PLGrid Infrastructure} through which access to the Prometheus supercomputer, located at ACK Cyfronet AGH in Krak\'{o}w, was provided. This work was supported in part by the Polish NCN grant 2019/33/B/ST9/01564 and the Black Hole Initiative at Harvard University, which is funded by grants the John Templeton Foundation and the Gordon and Betty Moore Foundation to Harvard University.
M.A. acknowledges the Polish NCN grant 2015/19/B/ST9/01099, while A.R. -- the NCN grant 2021/41/B/ST9/04110. M.A. and G.T. the Czech Science Foundation grant No. GX21-06825X, and D.L. the internal grant of SU SGS/13/2019 and the Student Grant Foundation of the Silesian University in Opava, Grant No. $\mathrm{SGF/1/2020}$, which has been carried out within the EU OPSRE project entitled ``Improving the quality of the internal grant scheme of the Silesian University in Opava'', reg. number: $\mathrm{CZ.02.2.69/0.0/0.0/19\_073/0016951}$. R.N. acknowledges the support of the NSF grants OISE-1743747 and AST-1816420. We also thank Alexandra Elbakyan for her contributions to the open science initiative. 

\section*{Data Availability}
The data underlying this article will be shared on reasonable request to the corresponding author.


\bibliographystyle{mnras}
\bibliography{bibliography}

\begin{thebibliography}{}
\makeatletter
\relax
\def\mn@urlcharsother{\let\do\@makeother \do\$\do\&\do\#\do\^\do\_\do\%\do\~}
\def\mn@doi{\begingroup\mn@urlcharsother \@ifnextchar [ {\mn@doi@}
  {\mn@doi@[]}}
\def\mn@doi@[#1]#2{\def\@tempa{#1}\ifx\@tempa\@empty \href
  {http://dx.doi.org/#2} {doi:#2}\else \href {http://dx.doi.org/#2} {#1}\fi
  \endgroup}
\def\mn@eprint#1#2{\mn@eprint@#1:#2::\@nil}
\def\mn@eprint@arXiv#1{\href {http://arxiv.org/abs/#1} {{\tt arXiv:#1}}}
\def\mn@eprint@dblp#1{\href {http://dblp.uni-trier.de/rec/bibtex/#1.xml}
  {dblp:#1}}
\def\mn@eprint@#1:#2:#3:#4\@nil{\def\@tempa {#1}\def\@tempb {#2}\def\@tempc
  {#3}\ifx \@tempc \@empty \let \@tempc \@tempb \let \@tempb \@tempa \fi \ifx
  \@tempb \@empty \def\@tempb {arXiv}\fi \@ifundefined
  {mn@eprint@\@tempb}{\@tempb:\@tempc}{\expandafter \expandafter \csname
  mn@eprint@\@tempb\endcsname \expandafter{\@tempc}}}

\bibitem[\protect\citeauthoryear{{Abramowicz} \& {Fragile}}{{Abramowicz} \&
  {Fragile}}{2013}]{accretion_review}
{Abramowicz} M.~A.,  {Fragile} P.~C.,  2013, \mn@doi [Living Reviews in
  Relativity] {10.12942/lrr-2013-1}, \href
  {https://ui.adsabs.harvard.edu/abs/2013LRR....16....1A} {16, 1}

\bibitem[\protect\citeauthoryear{{Abramowicz}, {Jaroszynski}  \&
  {Sikora}}{{Abramowicz} et~al.}{1978}]{Abramowicz1978}
{Abramowicz} M.,  {Jaroszynski} M.,   {Sikora} M.,  1978, \aap, \href
  {https://ui.adsabs.harvard.edu/abs/1978A&A....63..221A} {63, 221}

\bibitem[\protect\citeauthoryear{{Abramowicz}, {Czerny}, {Lasota}  \&
  {Szuszkiewicz}}{{Abramowicz} et~al.}{1988}]{Abramowicz1988}
{Abramowicz} M.~A.,  {Czerny} B.,  {Lasota} J.~P.,   {Szuszkiewicz} E.,  1988,
  \mn@doi [\apj] {10.1086/166683}, \href
  {https://ui.adsabs.harvard.edu/abs/1988ApJ...332..646A} {332, 646}

\bibitem[\protect\citeauthoryear{{Arnaud}}{{Arnaud}}{1996}]{Arnaud1996}
{Arnaud} K.~A.,  1996, in {Jacoby} G.~H.,  {Barnes} J.,  eds,  Astronomical
  Society of the Pacific Conference Series Vol. 101, Astronomical Data Analysis
  Software and Systems V. p.~17

\bibitem[\protect\citeauthoryear{{Bagi{\'n}ska}, {R{\'o}{\.z}a{\'n}ska},
  {Czerny}  \& {Janiuk}}{{Bagi{\'n}ska} et~al.}{2021}]{Baginska2021}
{Bagi{\'n}ska} P.,  {R{\'o}{\.z}a{\'n}ska} A.,  {Czerny} B.,   {Janiuk} A.,
  2021, \mn@doi [\apj] {10.3847/1538-4357/abee79}, \href
  {https://ui.adsabs.harvard.edu/abs/2021ApJ...912..110B} {912, 110}

\bibitem[\protect\citeauthoryear{{Begelman} \& {Pringle}}{{Begelman} \&
  {Pringle}}{2007}]{Begelman2007}
{Begelman} M.~C.,  {Pringle} J.~E.,  2007, \mn@doi [\mnras]
  {10.1111/j.1365-2966.2006.11372.x}, \href
  {https://ui.adsabs.harvard.edu/abs/2007MNRAS.375.1070B} {375, 1070}

\bibitem[\protect\citeauthoryear{{Ciesielski}, {Wielgus}, {Klu{\'z}niak}, {S{\k
  a}dowski}, {Abramowicz}, {Lasota}  \& {Rebusco}}{{Ciesielski}
  et~al.}{2012}]{Ciesielski2012}
{Ciesielski} A.,  {Wielgus} M.,  {Klu{\'z}niak} W.,  {S{\k a}dowski} A.,
  {Abramowicz} M.,  {Lasota} J.~P.,   {Rebusco} P.,  2012, \mn@doi [\aap]
  {10.1051/0004-6361/201117478}, \href
  {https://ui.adsabs.harvard.edu/abs/2012A&A...538A.148C} {538, A148}

\bibitem[\protect\citeauthoryear{{Davis} \& {Hubeny}}{{Davis} \&
  {Hubeny}}{2006}]{Davis2006}
{Davis} S.~W.,  {Hubeny} I.,  2006, \mn@doi [\apjs] {10.1086/503549}, \href
  {https://ui.adsabs.harvard.edu/abs/2006ApJS..164..530D} {164, 530}

\bibitem[\protect\citeauthoryear{{Dexter}, {Scepi}  \& {Begelman}}{{Dexter}
  et~al.}{2021}]{Dexter2021}
{Dexter} J.,  {Scepi} N.,   {Begelman} M.~C.,  2021, \mn@doi [\apjl]
  {10.3847/2041-8213/ac2608}, \href
  {https://ui.adsabs.harvard.edu/abs/2021ApJ...919L..20D} {919, L20}

\bibitem[\protect\citeauthoryear{{Done} \& {Kubota}}{{Done} \&
  {Kubota}}{2006}]{Done2006}
{Done} C.,  {Kubota} A.,  2006, \mn@doi [\mnras]
  {10.1111/j.1365-2966.2006.10737.x}, \href
  {https://ui.adsabs.harvard.edu/abs/2006MNRAS.371.1216D} {371, 1216}

\bibitem[\protect\citeauthoryear{{Done}, {Gierli{\'n}ski}  \& {Kubota}}{{Done}
  et~al.}{2007}]{Done2007}
{Done} C.,  {Gierli{\'n}ski} M.,   {Kubota} A.,  2007, \mn@doi [\aapr]
  {10.1007/s00159-007-0006-1}, \href
  {https://ui.adsabs.harvard.edu/abs/2007A&ARv..15....1D} {15, 1}

\bibitem[\protect\citeauthoryear{{EHT Collaboration} et~al.,}{{EHT
  Collaboration} et~al.}{2019}]{PaperI}
{EHT Collaboration} et~al., 2019, \mn@doi [\apjl] {10.3847/2041-8213/ab0ec7},
  \href {http://adsabs.harvard.edu/abs/2019ApJ...875L...1E} {875, L1}

\bibitem[\protect\citeauthoryear{{Giacconi}, {Kellogg}, {Gorenstein}, {Gursky}
  \& {Tananbaum}}{{Giacconi} et~al.}{1971}]{Giacconi1971}
{Giacconi} R.,  {Kellogg} E.,  {Gorenstein} P.,  {Gursky} H.,   {Tananbaum} H.,
   1971, \mn@doi [\apjl] {10.1086/180711}, \href
  {https://ui.adsabs.harvard.edu/abs/1971ApJ...165L..27G} {165, L27}

\bibitem[\protect\citeauthoryear{{Gierli{\'n}ski} \& {Done}}{{Gierli{\'n}ski}
  \& {Done}}{2004}]{Gierlinski2004}
{Gierli{\'n}ski} M.,  {Done} C.,  2004, \mn@doi [\mnras]
  {10.1111/j.1365-2966.2004.07266.x}, \href
  {https://ui.adsabs.harvard.edu/abs/2004MNRAS.347..885G} {347, 885}

\bibitem[\protect\citeauthoryear{{Gladstone}, {Roberts}  \& {Done}}{{Gladstone}
  et~al.}{2009}]{Gladstone2009}
{Gladstone} J.~C.,  {Roberts} T.~P.,   {Done} C.,  2009, \mn@doi [\mnras]
  {10.1111/j.1365-2966.2009.15123.x}, \href
  {https://ui.adsabs.harvard.edu/abs/2009MNRAS.397.1836G} {397, 1836}

\bibitem[\protect\citeauthoryear{{Gronkiewicz} \&
  {R{\'o}{\.z}a{\'n}ska}}{{Gronkiewicz} \&
  {R{\'o}{\.z}a{\'n}ska}}{2020}]{Gronki2020}
{Gronkiewicz} D.,  {R{\'o}{\.z}a{\'n}ska} A.,  2020, \mn@doi [\aap]
  {10.1051/0004-6361/201935033}, \href
  {https://ui.adsabs.harvard.edu/abs/2020A&A...633A..35G} {633, A35}

\bibitem[\protect\citeauthoryear{{Homan}, {Wijnands}, {van der Klis},
  {Belloni}, {van Paradijs}, {Klein-Wolt}, {Fender}  \& {M{\'e}ndez}}{{Homan}
  et~al.}{2001}]{Homan2001}
{Homan} J.,  {Wijnands} R.,  {van der Klis} M.,  {Belloni} T.,  {van Paradijs}
  J.,  {Klein-Wolt} M.,  {Fender} R.,   {M{\'e}ndez} M.,  2001, \mn@doi [\apjs]
  {10.1086/318954}, \href
  {https://ui.adsabs.harvard.edu/abs/2001ApJS..132..377H} {132, 377}

\bibitem[\protect\citeauthoryear{{Kaaret}, {Feng}  \& {Roberts}}{{Kaaret}
  et~al.}{2017}]{Kaaret2017}
{Kaaret} P.,  {Feng} H.,   {Roberts} T.~P.,  2017, \mn@doi [\araa]
  {10.1146/annurev-astro-091916-055259}, \href
  {https://ui.adsabs.harvard.edu/abs/2017ARA&A..55..303K} {55, 303}

\bibitem[\protect\citeauthoryear{{King}}{{King}}{2009}]{King2009}
{King} A.~R.,  2009, \mn@doi [\mnras] {10.1111/j.1745-3933.2008.00594.x}, \href
  {https://ui.adsabs.harvard.edu/abs/2009MNRAS.393L..41K} {393, L41}

\bibitem[\protect\citeauthoryear{{King}, {Davies}, {Ward}, {Fabbiano}  \&
  {Elvis}}{{King} et~al.}{2001}]{King2001}
{King} A.~R.,  {Davies} M.~B.,  {Ward} M.~J.,  {Fabbiano} G.,   {Elvis} M.,
  2001, \mn@doi [\apjl] {10.1086/320343}, \href
  {https://ui.adsabs.harvard.edu/abs/2001ApJ...552L.109K} {552, L109}

\bibitem[\protect\citeauthoryear{{Kluzniak} \& {Kita}}{{Kluzniak} \&
  {Kita}}{2000}]{Kluzniak2000}
{Kluzniak} W.,  {Kita} D.,  2000, arXiv e-prints, \href
  {https://ui.adsabs.harvard.edu/abs/2000astro.ph..6266K} {pp
  astro--ph/0006266}

\bibitem[\protect\citeauthoryear{{Kubota} \& {Makishima}}{{Kubota} \&
  {Makishima}}{2004}]{Kubota2004}
{Kubota} A.,  {Makishima} K.,  2004, \mn@doi [\apj] {10.1086/380433}, \href
  {https://ui.adsabs.harvard.edu/abs/2004ApJ...601..428K} {601, 428}

\bibitem[\protect\citeauthoryear{{Kulkarni} et~al.,}{{Kulkarni}
  et~al.}{2011}]{Kulkarni2011}
{Kulkarni} A.~K.,  et~al., 2011, \mn@doi [\mnras]
  {10.1111/j.1365-2966.2011.18446.x}, \href
  {https://ui.adsabs.harvard.edu/abs/2011MNRAS.414.1183K} {414, 1183}

\bibitem[\protect\citeauthoryear{{Lan{\v{c}}ov{\'a}}
  et~al.,}{{Lan{\v{c}}ov{\'a}} et~al.}{2019}]{Lancova2019}
{Lan{\v{c}}ov{\'a}} D.,  et~al., 2019, \mn@doi [\apjl]
  {10.3847/2041-8213/ab48f5}, \href
  {https://ui.adsabs.harvard.edu/abs/2019ApJ...884L..37L} {884, L37}

\bibitem[\protect\citeauthoryear{{Lasota}}{{Lasota}}{2001}]{Lasota2001}
{Lasota} J.-P.,  2001, \mn@doi [\nar] {10.1016/S1387-6473(01)00112-9}, \href
  {https://ui.adsabs.harvard.edu/abs/2001NewAR..45..449L} {45, 449}

\bibitem[\protect\citeauthoryear{{Lasota}, {Vieira}, {S{\k{a}}dowski},
  {Narayan}  \& {Abramowicz}}{{Lasota} et~al.}{2016}]{Lasota2016}
{Lasota} J.~P.,  {Vieira} R.~S.~S.,  {S{\k{a}}dowski} A.,  {Narayan} R.,
  {Abramowicz} M.~A.,  2016, \mn@doi [\aap] {10.1051/0004-6361/201527636},
  \href {https://ui.adsabs.harvard.edu/abs/2016A&A...587A..13L} {587, A13}

\bibitem[\protect\citeauthoryear{{Levine}, {Bradt}, {Cui}, {Jernigan},
  {Morgan}, {Remillard}, {Shirey}  \& {Smith}}{{Levine}
  et~al.}{1996}]{Levine1996}
{Levine} A.~M.,  {Bradt} H.,  {Cui} W.,  {Jernigan} J.~G.,  {Morgan} E.~H.,
  {Remillard} R.,  {Shirey} R.~E.,   {Smith} D.~A.,  1996, \mn@doi [\apjl]
  {10.1086/310260}, \href
  {https://ui.adsabs.harvard.edu/abs/1996ApJ...469L..33L} {469, L33}

\bibitem[\protect\citeauthoryear{{Li}, {Zimmerman}, {Narayan}  \&
  {McClintock}}{{Li} et~al.}{2005}]{Li2005}
{Li} L.-X.,  {Zimmerman} E.~R.,  {Narayan} R.,   {McClintock} J.~E.,  2005,
  \mn@doi [\apjs] {10.1086/428089}, \href
  {https://ui.adsabs.harvard.edu/abs/2005ApJS..157..335L} {157, 335}

\bibitem[\protect\citeauthoryear{{Lightman} \& {Eardley}}{{Lightman} \&
  {Eardley}}{1974}]{viscous74}
{Lightman} A.~P.,  {Eardley} D.~M.,  1974, \mn@doi [\apjl] {10.1086/181377},
  \href {https://ui.adsabs.harvard.edu/abs/1974ApJ...187L...1L} {187, L1}

\bibitem[\protect\citeauthoryear{{McClintock}, {Shafee}, {Narayan},
  {Remillard}, {Davis}  \& {Li}}{{McClintock} et~al.}{2006}]{McClintock2006}
{McClintock} J.~E.,  {Shafee} R.,  {Narayan} R.,  {Remillard} R.~A.,  {Davis}
  S.~W.,   {Li} L.-X.,  2006, \mn@doi [\apj] {10.1086/508457}, \href
  {https://ui.adsabs.harvard.edu/abs/2006ApJ...652..518M} {652, 518}

\bibitem[\protect\citeauthoryear{{McClintock}, {Narayan}  \&
  {Steiner}}{{McClintock} et~al.}{2014}]{McClintock2014}
{McClintock} J.~E.,  {Narayan} R.,   {Steiner} J.~F.,  2014, \mn@doi [\ssr]
  {10.1007/s11214-013-0003-9}, \href
  {https://ui.adsabs.harvard.edu/abs/2014SSRv..183..295M} {183, 295}

\bibitem[\protect\citeauthoryear{{M{\'e}ndez} \& {van der Klis}}{{M{\'e}ndez}
  \& {van der Klis}}{1997}]{Mendez1997}
{M{\'e}ndez} M.,  {van der Klis} M.,  1997, \mn@doi [\apj] {10.1086/303914},
  \href {https://ui.adsabs.harvard.edu/abs/1997ApJ...479..926M} {479, 926}

\bibitem[\protect\citeauthoryear{{Middleton}, {Done}, {Gierli{\'n}ski}  \&
  {Davis}}{{Middleton} et~al.}{2006}]{Middleton2006}
{Middleton} M.,  {Done} C.,  {Gierli{\'n}ski} M.,   {Davis} S.~W.,  2006,
  \mn@doi [\mnras] {10.1111/j.1365-2966.2006.11077.x}, \href
  {https://ui.adsabs.harvard.edu/abs/2006MNRAS.373.1004M} {373, 1004}

\bibitem[\protect\citeauthoryear{{Mishra}, {Begelman}, {Armitage}  \&
  {Simon}}{{Mishra} et~al.}{2020}]{Mishra2020}
{Mishra} B.,  {Begelman} M.~C.,  {Armitage} P.~J.,   {Simon} J.~B.,  2020,
  \mn@doi [\mnras] {10.1093/mnras/stz3572}, \href
  {https://ui.adsabs.harvard.edu/abs/2020MNRAS.492.1855M} {492, 1855}

\bibitem[\protect\citeauthoryear{{Mondal}, {R{\'o}{\.z}a{\'n}ska},
  {Bagi{\'n}ska}, {Markowitz}  \& {De Marco}}{{Mondal}
  et~al.}{2021}]{Mondal2021}
{Mondal} S.,  {R{\'o}{\.z}a{\'n}ska} A.,  {Bagi{\'n}ska} P.,  {Markowitz} A.,
  {De Marco} B.,  2021, \mn@doi [\aap] {10.1051/0004-6361/202140459}, \href
  {https://ui.adsabs.harvard.edu/abs/2021A&A...651A..54M} {651, A54}

\bibitem[\protect\citeauthoryear{{Motta}, {Kajava},
  {S{\'a}nchez-Fern{\'a}ndez}, {Giustini}  \& {Kuulkers}}{{Motta}
  et~al.}{2017}]{Motta2017}
{Motta} S.~E.,  {Kajava} J.~J.~E.,  {S{\'a}nchez-Fern{\'a}ndez} C.,  {Giustini}
  M.,   {Kuulkers} E.,  2017, \mn@doi [\mnras] {10.1093/mnras/stx466}, \href
  {https://ui.adsabs.harvard.edu/abs/2017MNRAS.468..981M} {468, 981}

\bibitem[\protect\citeauthoryear{{Narayan} \& {McClintock}}{{Narayan} \&
  {McClintock}}{2005}]{Narayan2005}
{Narayan} R.,  {McClintock} J.~E.,  2005, \mn@doi [\apj] {10.1086/428709},
  \href {https://ui.adsabs.harvard.edu/abs/2005ApJ...623.1017N} {623, 1017}

\bibitem[\protect\citeauthoryear{{Narayan}, {Zhu}, {Psaltis}  \& {S{\k
  a}dowski}}{{Narayan} et~al.}{2016}]{Narayan2016}
{Narayan} R.,  {Zhu} Y.,  {Psaltis} D.,   {S{\k a}dowski} A.,  2016, \mn@doi
  [\mnras] {10.1093/mnras/stv2979}, \href
  {https://ui.adsabs.harvard.edu/abs/2016MNRAS.457..608N} {457, 608}

\bibitem[\protect\citeauthoryear{{Narayan}, {Sa{\k{a}}dowski}  \&
  {Soria}}{{Narayan} et~al.}{2017}]{Narayan2017}
{Narayan} R.,  {Sa{\k{a}}dowski} A.,   {Soria} R.,  2017, \mn@doi [\mnras]
  {10.1093/mnras/stx1027}, \href
  {https://ui.adsabs.harvard.edu/abs/2017MNRAS.469.2997N} {469, 2997}

\bibitem[\protect\citeauthoryear{{Novikov} \& {Thorne}}{{Novikov} \&
  {Thorne}}{1973}]{Novikov1973}
{Novikov} I.~D.,  {Thorne} K.~S.,  1973, in {Dewitt} C.,  {Dewitt} B.~S.,  eds,
  Black Holes (Les Astres Occlus). pp 343--450

\bibitem[\protect\citeauthoryear{{Oda}, {Machida}, {Nakamura}  \&
  {Matsumoto}}{{Oda} et~al.}{2009}]{Oda2009}
{Oda} H.,  {Machida} M.,  {Nakamura} K.~E.,   {Matsumoto} R.,  2009, \mn@doi
  [\apj] {10.1088/0004-637X/697/1/16}, \href
  {https://ui.adsabs.harvard.edu/abs/2009ApJ...697...16O} {697, 16}

\bibitem[\protect\citeauthoryear{{Ogawa}, {Ohsuga}, {Makino}  \&
  {Mineshige}}{{Ogawa} et~al.}{2021}]{Ogawa2021}
{Ogawa} T.,  {Ohsuga} K.,  {Makino} Y.,   {Mineshige} S.,  2021, \mn@doi
  [\pasj] {10.1093/pasj/psab031}, \href
  {https://ui.adsabs.harvard.edu/abs/2021PASJ...73..701O} {73, 701}

\bibitem[\protect\citeauthoryear{{Ohsuga} \& {Mineshige}}{{Ohsuga} \&
  {Mineshige}}{2011}]{Ohsuga2011}
{Ohsuga} K.,  {Mineshige} S.,  2011, \mn@doi [\apj]
  {10.1088/0004-637X/736/1/2}, \href
  {https://ui.adsabs.harvard.edu/abs/2011ApJ...736....2O} {736, 2}

\bibitem[\protect\citeauthoryear{{Ohsuga}, {Mineshige}, {Mori}  \&
  {Kato}}{{Ohsuga} et~al.}{2009}]{Ohsuga2009}
{Ohsuga} K.,  {Mineshige} S.,  {Mori} M.,   {Kato} Y.,  2009, \mn@doi [\pasj]
  {10.1093/pasj/61.3.L7}, \href
  {https://ui.adsabs.harvard.edu/abs/2009PASJ...61L...7O} {61, L7}

\bibitem[\protect\citeauthoryear{{Page} \& {Thorne}}{{Page} \&
  {Thorne}}{1974}]{Page1974}
{Page} D.~N.,  {Thorne} K.~S.,  1974, \mn@doi [\apj] {10.1086/152990}, \href
  {https://ui.adsabs.harvard.edu/abs/1974ApJ...191..499P} {191, 499}

\bibitem[\protect\citeauthoryear{{Regev} \& {Gitelman}}{{Regev} \&
  {Gitelman}}{2002}]{Regev2002}
{Regev} O.,  {Gitelman} L.,  2002, \mn@doi [\aap] {10.1051/0004-6361:20021492},
  \href {https://ui.adsabs.harvard.edu/abs/2002A&A...396..623R} {396, 623}

\bibitem[\protect\citeauthoryear{{Remillard} \& {McClintock}}{{Remillard} \&
  {McClintock}}{2006}]{Remillard2006}
{Remillard} R.~A.,  {McClintock} J.~E.,  2006, \mn@doi [\araa]
  {10.1146/annurev.astro.44.051905.092532}, \href
  {https://ui.adsabs.harvard.edu/abs/2006ARA&A..44...49R} {44, 49}

\bibitem[\protect\citeauthoryear{{R{\'o}{\.z}a{\'n}ska}, {Czerny}, {{\.Z}ycki}
  \& {Pojma{\'n}ski}}{{R{\'o}{\.z}a{\'n}ska} et~al.}{1999}]{Rozanska1999}
{R{\'o}{\.z}a{\'n}ska} A.,  {Czerny} B.,  {{\.Z}ycki} P.~T.,   {Pojma{\'n}ski}
  G.,  1999, \mn@doi [\mnras] {10.1046/j.1365-8711.1999.02425.x}, \href
  {https://ui.adsabs.harvard.edu/abs/1999MNRAS.305..481R} {305, 481}

\bibitem[\protect\citeauthoryear{{R{\'o}{\.z}a{\'n}ska}, {Madej}, {Konorski}
  \& {S{\k{a}}dowski}}{{R{\'o}{\.z}a{\'n}ska} et~al.}{2011}]{Rozanska2011}
{R{\'o}{\.z}a{\'n}ska} A.,  {Madej} J.,  {Konorski} P.,   {S{\k{a}}dowski} A.,
  2011, \mn@doi [\aap] {10.1051/0004-6361/201015626}, \href
  {https://ui.adsabs.harvard.edu/abs/2011A&A...527A..47R} {527, A47}

\bibitem[\protect\citeauthoryear{{R{\'o}{\.z}a{\'n}ska}, {Malzac}, {Belmont},
  {Czerny}  \& {Petrucci}}{{R{\'o}{\.z}a{\'n}ska} et~al.}{2015}]{Rozanska2015}
{R{\'o}{\.z}a{\'n}ska} A.,  {Malzac} J.,  {Belmont} R.,  {Czerny} B.,
  {Petrucci} P.~O.,  2015, \mn@doi [\aap] {10.1051/0004-6361/201526288}, \href
  {https://ui.adsabs.harvard.edu/abs/2015A&A...580A..77R} {580, A77}

\bibitem[\protect\citeauthoryear{{Shakura} \& {Sunyaev}}{{Shakura} \&
  {Sunyaev}}{1973}]{Shakura1973}
{Shakura} N.~I.,  {Sunyaev} R.~A.,  1973, \aap, \href
  {https://ui.adsabs.harvard.edu/abs/1973A&A....24..337S} {500, 33}

\bibitem[\protect\citeauthoryear{{Shakura} \& {Sunyaev}}{{Shakura} \&
  {Sunyaev}}{1976}]{thermal76}
{Shakura} N.~I.,  {Sunyaev} R.~A.,  1976, \mn@doi [\mnras]
  {10.1093/mnras/175.3.613}, \href
  {https://ui.adsabs.harvard.edu/abs/1976MNRAS.175..613S} {175, 613}

\bibitem[\protect\citeauthoryear{{Sikora}}{{Sikora}}{1981}]{Sikora1981}
{Sikora} M.,  1981, \mn@doi [\mnras] {10.1093/mnras/196.2.257}, \href
  {https://ui.adsabs.harvard.edu/abs/1981MNRAS.196..257S} {196, 257}

\bibitem[\protect\citeauthoryear{{S{\k a}dowski}}{{S{\k
  a}dowski}}{2016}]{Sadowski2016}
{S{\k a}dowski} A.,  2016, \mn@doi [\mnras] {10.1093/mnras/stw913}, \href
  {https://ui.adsabs.harvard.edu/abs/2016MNRAS.459.4397S} {459, 4397}

\bibitem[\protect\citeauthoryear{{S{\k a}dowski}, {Narayan}, {Tchekhovskoy}  \&
  {Zhu}}{{S{\k a}dowski} et~al.}{2013}]{Sadowski2013}
{S{\k a}dowski} A.,  {Narayan} R.,  {Tchekhovskoy} A.,   {Zhu} Y.,  2013,
  \mn@doi [\mnras] {10.1093/mnras/sts632}, \href
  {https://ui.adsabs.harvard.edu/abs/2013MNRAS.429.3533S} {429, 3533}

\bibitem[\protect\citeauthoryear{{S{\k a}dowski}, {Narayan}, {McKinney}  \&
  {Tchekhovskoy}}{{S{\k a}dowski} et~al.}{2014}]{Sadowski2014}
{S{\k a}dowski} A.,  {Narayan} R.,  {McKinney} J.~C.,   {Tchekhovskoy} A.,
  2014, \mn@doi [\mnras] {10.1093/mnras/stt2479}, \href
  {https://ui.adsabs.harvard.edu/abs/2014MNRAS.439..503S} {439, 503}

\bibitem[\protect\citeauthoryear{{S{\k{a}}dowski}}{{S{\k{a}}dowski}}{2009}]{Sadowski2009}
{S{\k{a}}dowski} A.,  2009, \mn@doi [\apjs] {10.1088/0067-0049/183/2/171},
  \href {https://ui.adsabs.harvard.edu/abs/2009ApJS..183..171S} {183, 171}

\bibitem[\protect\citeauthoryear{{S{\k{a}}dowski} \&
  {Narayan}}{{S{\k{a}}dowski} \& {Narayan}}{2015}]{Sadowski2015}
{S{\k{a}}dowski} A.,  {Narayan} R.,  2015, \mn@doi [\mnras]
  {10.1093/mnras/stv2022}, \href
  {https://ui.adsabs.harvard.edu/abs/2015MNRAS.454.2372S} {454, 2372}

\bibitem[\protect\citeauthoryear{{S{\k{a}}dowski}, {Abramowicz}, {Bursa},
  {Klu{\'z}niak}, {Lasota}  \& {R{\'o}{\.z}a{\'n}ska}}{{S{\k{a}}dowski}
  et~al.}{2011}]{Sadowski2011}
{S{\k{a}}dowski} A.,  {Abramowicz} M.,  {Bursa} M.,  {Klu{\'z}niak} W.,
  {Lasota} J.~P.,   {R{\'o}{\.z}a{\'n}ska} A.,  2011, \mn@doi [\aap]
  {10.1051/0004-6361/201015256}, \href
  {https://ui.adsabs.harvard.edu/abs/2011A&A...527A..17S} {527, A17}

\bibitem[\protect\citeauthoryear{{S{\k{a}}dowski}, {Wielgus}, {Narayan},
  {Abarca}, {McKinney}  \& {Chael}}{{S{\k{a}}dowski}
  et~al.}{2017}]{Sadowski2017}
{S{\k{a}}dowski} A.,  {Wielgus} M.,  {Narayan} R.,  {Abarca} D.,  {McKinney}
  J.~C.,   {Chael} A.,  2017, \mn@doi [\mnras] {10.1093/mnras/stw3116}, \href
  {https://ui.adsabs.harvard.edu/abs/2017MNRAS.466..705S} {466, 705}

\bibitem[\protect\citeauthoryear{{Steiner}, {McClintock}, {Remillard},
  {Narayan}  \& {Gou}}{{Steiner} et~al.}{2009}]{Steiner2009}
{Steiner} J.~F.,  {McClintock} J.~E.,  {Remillard} R.~A.,  {Narayan} R.,
  {Gou} L.,  2009, \mn@doi [\apjl] {10.1088/0004-637X/701/2/L83}, \href
  {https://ui.adsabs.harvard.edu/abs/2009ApJ...701L..83S} {701, L83}

\bibitem[\protect\citeauthoryear{{Steiner}, {McClintock}, {Orosz}, {Remillard},
  {Bailyn}, {Kolehmainen}  \& {Straub}}{{Steiner} et~al.}{2014}]{Steiner2014}
{Steiner} J.~F.,  {McClintock} J.~E.,  {Orosz} J.~A.,  {Remillard} R.~A.,
  {Bailyn} C.~D.,  {Kolehmainen} M.,   {Straub} O.,  2014, \mn@doi [\apjl]
  {10.1088/2041-8205/793/2/L29}, \href
  {https://ui.adsabs.harvard.edu/abs/2014ApJ...793L..29S} {793, L29}

\bibitem[\protect\citeauthoryear{{Straub} et~al.,}{{Straub}
  et~al.}{2011}]{Straub2011}
{Straub} O.,  et~al., 2011, \mn@doi [\aap] {10.1051/0004-6361/201117385}, \href
  {https://ui.adsabs.harvard.edu/abs/2011A&A...533A..67S} {533, A67}

\bibitem[\protect\citeauthoryear{{Sutton}, {Roberts}  \& {Middleton}}{{Sutton}
  et~al.}{2013}]{Sutton2013}
{Sutton} A.~D.,  {Roberts} T.~P.,   {Middleton} M.~J.,  2013, \mn@doi [\mnras]
  {10.1093/mnras/stt1419}, \href
  {https://ui.adsabs.harvard.edu/abs/2013MNRAS.435.1758S} {435, 1758}

\bibitem[\protect\citeauthoryear{Vincent, Paumard, Gourgoulhon  \&
  Perrin}{Vincent et~al.}{2011}]{Vincent2011}
Vincent F.~H.,  Paumard T.,  Gourgoulhon E.,   Perrin G.,  2011, \mn@doi
  [Classical and Quantum Gravity] {10.1088/0264-9381/28/22/225011}, 28, 225011

\bibitem[\protect\citeauthoryear{{Wielgus}, {Yan}, {Lasota}  \&
  {Abramowicz}}{{Wielgus} et~al.}{2016}]{Wielgus2016}
{Wielgus} M.,  {Yan} W.,  {Lasota} J.~P.,   {Abramowicz} M.~A.,  2016, \mn@doi
  [\aap] {10.1051/0004-6361/201527878}, \href
  {https://ui.adsabs.harvard.edu/abs/2016A&A...587A..38W} {587, A38}

\bibitem[\protect\citeauthoryear{{Zdziarski}, {Johnson}  \&
  {Magdziarz}}{{Zdziarski} et~al.}{1996}]{Zdziarski1996}
{Zdziarski} A.~A.,  {Johnson} W.~N.,   {Magdziarz} P.,  1996, \mn@doi [\mnras]
  {10.1093/mnras/283.1.193}, \href
  {https://ui.adsabs.harvard.edu/abs/1996MNRAS.283..193Z} {283, 193}

\bibitem[\protect\citeauthoryear{{Zhang}, {Cui}, {Chen}, {Yao}, {Zhang}, {Sun},
  {Wu}  \& {Xu}}{{Zhang} et~al.}{2000}]{Zhang2000}
{Zhang} S.~N.,  {Cui} W.,  {Chen} W.,  {Yao} Y.,  {Zhang} X.,  {Sun} X.,  {Wu}
  X.-B.,   {Xu} H.,  2000, \mn@doi [Science] {10.1126/science.287.5456.1239},
  \href {https://ui.adsabs.harvard.edu/abs/2000Sci...287.1239Z} {287, 1239}

\bibitem[\protect\citeauthoryear{{Zhu} \& {Narayan}}{{Zhu} \&
  {Narayan}}{2013}]{Zhu2013}
{Zhu} Y.,  {Narayan} R.,  2013, \mn@doi [\mnras] {10.1093/mnras/stt1161}, \href
  {https://ui.adsabs.harvard.edu/abs/2013MNRAS.434.2262Z} {434, 2262}

\bibitem[\protect\citeauthoryear{{Zhu}, {Narayan}, {S{\k a}dowski}  \&
  {Psaltis}}{{Zhu} et~al.}{2015}]{Zhu2015}
{Zhu} Y.,  {Narayan} R.,  {S{\k a}dowski} A.,   {Psaltis} D.,  2015, \mn@doi
  [\mnras] {10.1093/mnras/stv1046}, \href
  {https://ui.adsabs.harvard.edu/abs/2015MNRAS.451.1661Z} {451, 1661}

\bibitem[\protect\citeauthoryear{{{\.Z}ycki}, {Done}  \& {Smith}}{{{\.Z}ycki}
  et~al.}{1999}]{Zycki1999}
{{\.Z}ycki} P.~T.,  {Done} C.,   {Smith} D.~A.,  1999, \mn@doi [\mnras]
  {10.1046/j.1365-8711.1999.02885.x}, \href
  {https://ui.adsabs.harvard.edu/abs/1999MNRAS.309..561Z} {309, 561}

\makeatother
\end{thebibliography}

\bsp	
\label{lastpage}
\end{document}